\documentclass[prb,aps,twocolumn,superscriptaddress,floatfix,citeautoscript,showpacs]{revtex4}
\usepackage{graphicx,rotating,subfigure,amsmath,amsfonts,amssymb,delarray,color}
\usepackage{dsfont}
\usepackage[T1]{fontenc}
\usepackage{multirow}
\usepackage{comment}

\newcommand{\im}{\text{i}}

\newcommand{\tr}{\textrm{tr}}

\def\12{\frac{1}{2}}

\begin{document}
\title{Current reversals and metastable states in the infinite\\ Bose-Hubbard chain with local particle loss}
\author{M.~Kiefer-Emmanouilidis}
\affiliation{University of Manitoba, Department of Physics, Winnipeg}
\affiliation{Technische Universit\"at Kaiserslautern, Department of Physics, Kaiserslautern}
\author{J.~Sirker}
\affiliation{University of Manitoba, Department of Physics, Winnipeg}
\date{\today}

\begin{abstract}
We present an algorithm which combines the quantum trajectory approach
to open quantum systems with a density-matrix renormalization group
scheme for infinite one-dimensional lattice systems. We apply this
method to investigate the long-time dynamics in the Bose-Hubbard model
with local particle loss starting from a Mott-insulating initial state
with one boson per site. While the short-time dynamics can be
described even quantitatively by an equation of motion (EOM) approach
at the mean-field level, many-body interactions lead to unexpected
effects at intermediate and long times: local particle currents far
away from the dissipative site start to reverse direction ultimately
leading to a metastable state with a total particle current pointing
away from the lossy site. An alternative EOM approach based on an
effective fermion model shows that the reversal of currents can be
understood qualitatively by the creation of holon-doublon pairs at the
edge of the region of reduced particle density. The doublons are then
able to escape while the holes move towards the dissipative site, a
process reminiscent---in a loose sense---of Hawking radiation.
\end{abstract}

\maketitle

\section{Introduction}
No quantum system is perfectly isolated. Coherent dynamics as
described by Schr\" odinger's equation lasts only over a finite
timescale before dissipation leads to decoherence. While dissipation
is an intrinsic process in solid state systems determined by the
properties of the material, the advent of quantum gases in optical
lattices \cite{Bloch_NatPhys} has made it possible to study lattice
systems where dissipation can be controlled to a certain degree and
used as a tool to manipulate the quantum
state.\cite{TomitaNakajima,Daley2014,Verstraete2009b}

Experimentally it has been shown, for example, that strong dissipation
in the form of two-body losses can model a Pauli exclusion principle,
fermionizing a system.\cite{Syassen2008} Using an electron beam, a
controlled local particle loss process has been realized for a
Bose-Einstein condensate (BEC) providing direct evidence for the
quantum Zeno effect.\cite{BarontiniLabouvie} Furthermore, local
particle loss has been used to create a tunnel junction between two
Bose-Einstein condensates (BEC) and negative differential conductance
has been observed.\cite{Labouvie2015a} For a one-dimensional array of
BEC's with a single lossy site it has also been shown that a
transition from a superfluid to a resistive state can be driven by
tuning the loss rate $\gamma$ with a bistability occuring at
intermediate $\gamma$.\cite{Labouvie2016}

Theoretically, local particle loss in the non-interacting Bose-Hubbard
model has been studied in Ref.~\onlinecite{Kepesidis2012} while the
interacting case has been investigated numerically using
time-dependent density-matrix renormalization group (tDMRG)
algorithms.\cite{Barmettler2011} The quantum Zeno dynamics which has
been observed in these simulations for local particle loss rates
$\gamma$ much larger than the hopping amplitude $J$ can be understood
in a perturbative approach based on adiabatic
elimination.\cite{Garcia-Ripoll2009} Global three-body loss processes
have also been simulated by tDMRG algorithms and have been shown to
give rise to effective three-body hard-core
interactions.\cite{Daley2009} Quite recently, also the cases of
interacting spinless fermions with disorder and local particle
loss\cite{NieuwenburgMalo} and of the Bose-Hubbard model with
dephasing have been studied.\cite{BernierTan}

In this paper we will consider open quantum systems which can be
described in Markov approximation leading to the following general
Lindblad master equation (LME) for the density matrix $\hat\rho$
\begin{equation}
\label{LME1}
\frac{d}{dt}\hat{\rho} = -\mathrm{i}[\hat{H},\hat{\rho}] + \sum\limits_{j=1}^{L} \gamma_j\left(\hat{A}_j\hat{\rho} \hat{A}_j^\dagger - \frac{1}{2}\left\{\hat{A}_j^\dagger \hat{A}_j, \hat{\rho}\right\}\right).
\end{equation}
Here $H$ is the Hamiltonian, $\hat A_j$ the operator
describing local dissipation at site $j$ of a lattice of length
$L$ with rate $\gamma_j$, and $\{.,.\}$ the anti-commutator.
\begin{figure}[!ht]
\includegraphics[width=0.98\linewidth]{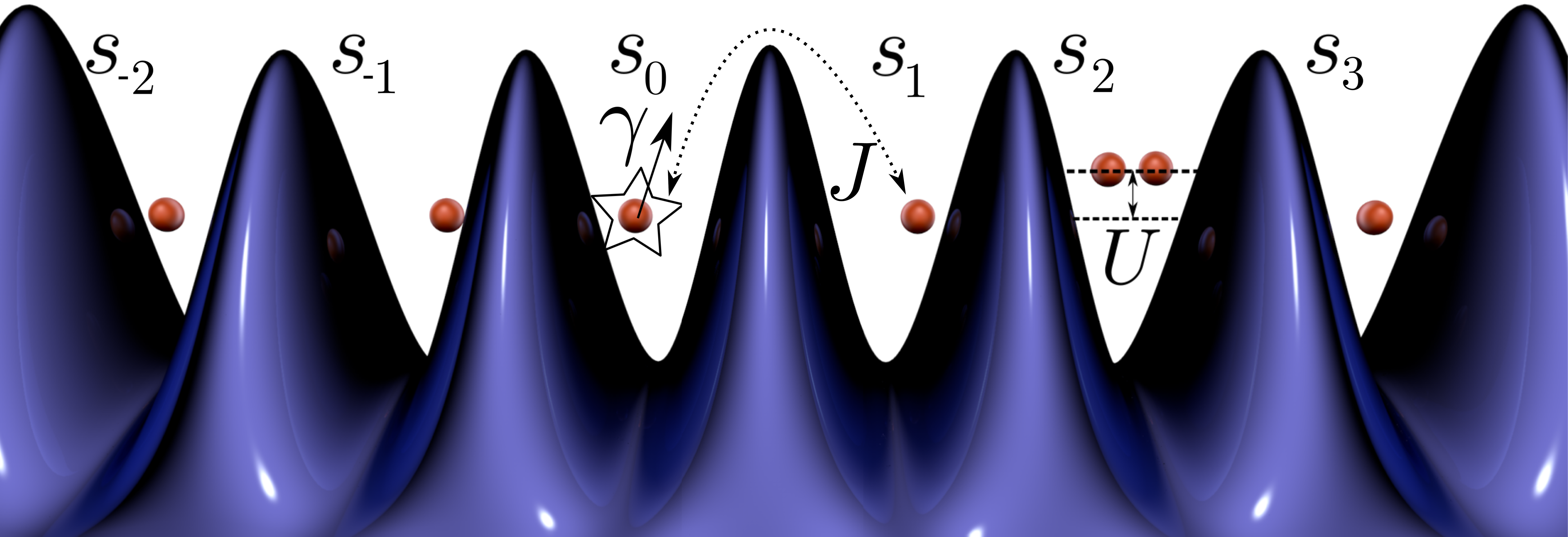}
\caption{A one-dimensional lattice model with hopping amplitude $J$ and onsite interaction $U$. 
At site $j=0$ particles escape the lattice with loss rate $\gamma$.}
\label{OpenBosehub}
\end{figure}

Part of the progress in studying the dynamics of one-dimensional open
many-body systems is currently driven by numerical renormalisation
group algorithms such as tDMRG
\cite{White1992,DaleyKollath,WhiteFeiguin} and the time-evolving
block decimation (TEBD)
\cite{Vidal2003,Zwolak2004,PhysRevLett.93.207204} for finite lattice
systems. For local particle loss neither method unfortunately leads to
non-equilibrium steady states (NESS) other than the vacuum because the
number of particles is also typically finite. The same problem also
exists for metastable states established at long time
scales.\cite{Macieszczak2016} Here we present a numerical scheme
combining the quantum trajectory (QT) approach
\cite{Molmer1993,Carmichael1993,Daley2014,Theoopenqusys2002} with the Light Cone 
Renormalization Group (LCRG)
\cite{EnssSirker} to treat open one-dimensional quantum systems
directly in the thermodynamic limit. This will allow, in particular,
to study the dynamics in the Bose-Hubbard model with local particle
loss shown in Fig.~\ref{OpenBosehub} at times $t\gg J/\gamma$.

Our paper is organized as follows. In Sec.~\ref{Model} we introduce
the Bose-Hubbard chain with local particle loss. We then discuss
equation of motion (EOM) approaches in Sec.~\ref{Sec_EOM} before
describing the numerical renormalization group algorithm to simulate
the Lindblad dynamics for infinite system size in
Sec.~\ref{Sec_LCRG}. The results of both methods are presented in
Sec.~\ref{Results} which includes a discussion of the density and
current profiles, the particle loss rate, and the evolution of the
density-density correlations. Sec.~\ref{Concl} is devoted to a short
summary and conclusions.

\section{Model and Methods}
\label{Model}
In the following, we will consider the Bose-Hubbard Hamiltonian 
\begin{equation}
H = -J\sum_j \left( \hat{b}^\dagger_j \hat{b}_{j+1} + h.c. \right) + \frac{U}{2} \sum\limits _j \hat{n}_j\left( \hat{n}_j - 1 \right) - \mu \sum \limits_j \hat{n}_j,
\end{equation}
where $\hat{b}_j^{(\dagger)}$ is the bosonic annihilation (creation)
operator acting at site $j$, and $\hat{n}_j = \hat{b}_j^\dagger
\hat{b}_j$ is the number operator. The bosonic operators fulfill
the commutation relations $[\hat{b}_i,\hat{b}_j^\dagger] =
\delta_{ij}$ and $[\hat{b}_i^\dagger,\hat{b}_j^\dagger] =
[\hat{b}_i,\hat{b}_j]=0$, where $\delta_{ij}$ is the Kronecker
delta. $J$ is the hopping amplitude and $U$ the onsite Hubbard
interaction which is assumed to be positive corresponding to repulsive
interactions between atoms on the same site. $\mu$ is the chemical
potential. We assume that the system at time $t=0$ is prepared in the
ground state of the closed system and concentrate, in particular, on
initial states with commensurate filling $\langle n_j\rangle=1$ deep
in the Mott insulating phase ($U\gg 3J$). The non-unitary dynamics is
then described by the LME
\begin{equation}
\label{LME2}
\frac{d}{dt}\hat{\rho} = -\mathrm{i}[\hat{H}_{\mathrm{BH}},\hat{\rho}] +  \gamma\left(\hat{b}_0\hat{\rho} \hat{b}_0^\dagger - \frac{1}{2}\left\{\hat{b}_0^\dagger \hat{b}_0, \hat{\rho}\right\}\right),
\end{equation}
which is a special case of the general LME, Eq.~\eqref{LME1}, with
dissipation---in terms of a local particle loss process---limited to
site $j=0$. The model is motivated by recent experiments on cold
atomic gases where an electron beam has been used to ionize and eject
particles from the gas with single site
resolution.\cite{WuertzLangen,Labouvie2015a, Labouvie2016}

\subsection{Equation of motion}
\label{Sec_EOM}
The time-dependence of an observable $\hat X(t)$ in an open quantum
system modeled by an LME is given by the EOM
\begin{equation}
\label{EOM}
\frac{d}{dt} \hat X = \im [\hat H,\hat X] + \sum\limits_{j=1}^{L} \gamma_j \left(\hat{A}^\dagger_j\hat X\hat{A}_j - \frac{1}{2}\left\{\hat{A}_i^\dagger \hat{A}_i, \hat{X}\right\}\right).
\end{equation} 
For particle loss $\hat A_j = \hat b_j$ and without the Hubbard
interaction $U$ the EOM closes and the dynamics can be obtained
exactly by numerically integrating the EOM. For finite interactions,
on the other hand, terms will in general be generated on the right
hand side of Eq.~\eqref{EOM} which contain more bosonic operators than
the observable $\hat X$ leading to an infinite hierarchy of
equations. This hierarchy has to be truncated in practice by using a
mean-field decoupling of higher order correlators. Nevertheless, for
short times such an approach often yields a good approximation of the
non-equilibrium dynamics of local observables.

\subsubsection{Direct decoupling}
\label{Decoupling}
We are interested, in particular, in the time evolution of the density
profiles $\langle n_j\rangle(t)$ and current profiles $\langle
\mathcal{J}_j\rangle(t)$. Evaluating Eq.~\eqref{EOM} for the two-point function
$\sigma_{jk}(t)=\langle b_j^\dagger b_k\rangle(t)$ leads
to\cite{WimbergerReview,KordasWitthaut}
\begin{eqnarray}
\label{EOM2}
\im\frac{d}{dt}\sigma_{jk} &=& -J\left(\sigma_{j,k+1}+\sigma_{j,k-1}-\sigma_{j+1,k}-\sigma_{j-1,k}\right) \nonumber \\
&+& U\left(\langle a_j^\dagger a_k^\dagger a_k a_k\rangle -\langle a_j^\dagger a_j^\dagger a_j a_k\rangle \right) \nonumber \\
&-&\im\frac{\gamma_0}{2}\left(\delta_{j,0}+\delta_{k,0}\right)\sigma_{jk}.	
\end{eqnarray}
In a first order approximation, we can simply use a Hartree-Fock
decoupling of the quartic terms
\begin{equation}
\label{EOM2p1}
\langle a_j^\dagger a_k^\dagger a_k
a_k\rangle -\langle a_j^\dagger a_j^\dagger a_j a_k\rangle\to
\sigma_{kk}\sigma_{jk}-\sigma_{jj}\sigma_{jk}\, . 
\end{equation}
Within this decoupling scheme, Eq.~\eqref{EOM2} can now be solved
numerically. To improve on this approximation and to check how
sensitive the solution is to the decoupling, we also consider the EOM
for a general four-point correlator $\theta_{ijkl}=\langle b_i^\dagger
b_j^\dagger b_k b_l\rangle$:
\begin{eqnarray}
\label{EOM3}
&&\im\frac{d}{dt}\theta_{ijkl} = -J\left(\theta_{i,j,k-1,l}+\theta_{i,j,k+1,l}+\theta_{i,j,k,l-1}\right. \nonumber \\
&+& \left.\theta_{i,j,k,l+1}-\theta_{i-1,j,k,l}-\theta_{i+1,j,k,l}-\theta_{i,j-1,k,l}-\theta_{i,j+1,k,l}\right) \nonumber \\
&+& U \left( \langle b_i^\dagger b_j^\dagger b_j^\dagger b_j b_k b_l\rangle + \langle b_i^\dagger b_i^\dagger b_i b_j^\dagger b_k b_l\rangle\right. \nonumber \\
&-& \left. \langle b_i^\dagger b_j^\dagger b_k b_l^\dagger b_l b_l\rangle - \langle b_i^\dagger b_j^\dagger b_k^\dagger b_l b_k b_k\rangle\right) \nonumber \\
&-& \im\frac{\gamma_0}{2}\left(\delta_{i,0}+\delta_{j,0}+\delta_{k,0}+\delta_{l,0}\right)\theta_{ijkl}.	
\end{eqnarray}
To close the system of EOM's, Eqs.~(\ref{EOM2},\ref{EOM3}), we now decouple
the six-point correlators into 4-point and two-point correlators
\begin{eqnarray}
\label{EOM3p1}
&& U \left( \langle b_i^\dagger b_j^\dagger b_j^\dagger b_j b_k b_l\rangle + \langle b_i^\dagger b_i^\dagger b_i b_j^\dagger b_k b_l\rangle\right. \nonumber \\
&& - \left.\langle b_i^\dagger b_j^\dagger b_k b_l^\dagger b_l b_l\rangle - \langle b_i^\dagger b_j^\dagger b_k^\dagger b_l b_k b_k\rangle\right) \nonumber \\
&& \to U\theta_{ijkl}\left(\sigma_{jj} +\sigma_{ii}-\sigma_{kk}-\sigma_{ll}\right).
\end{eqnarray}
In the following, we denote the EOM \eqref{EOM2} with the Hartree-Fock
decoupling \eqref{EOM2p1} as first order approximation and the EOM's
(\ref{EOM2},\ref{EOM3}) with the decoupling scheme \eqref{EOM3p1} as
second order approximation.

\subsubsection{Effective fermionic model}
\label{Sec_EFM}
Alternatively, an EOM approach can be formulated by first mapping the
BHM for strong repulsive interactions onto an effective fermionic
model
(EFM).\cite{BarmettlerPoletti,CheneauBarmettler,AndraschkoSirker2,BernierTan}
The main idea is to limit the local Hilbert space to states with
$n=0,1,2$ particles. We can then interpret the state $|1\rangle$ as
the vacuum, the holon as a fermion with spin down, and the doublon as
a fermion with spin up. The fermionic statistics ensures that not more
than holon or doublon can occupy the same site. Formally, the mapping
is given by
\begin{equation}
\label{EFM1}
b_j^\dagger = Z_j \sqrt{2}c_{j\uparrow}^\dagger (1-n_{j\downarrow})+Z_j c_{j\downarrow} (1-n_{j\uparrow})
\end{equation}
with $n_{j\sigma}=c^\dagger_{j\sigma}c_{j\sigma}$ and the
Jordan-Wigner string $Z_j=\prod_{j'<j}\exp(\im\pi\sum_\sigma
n_{j'\sigma})$. The local density operator then reads
\begin{equation}
\label{EFM2}
b_j^\dagger b_j = 1 + n_{j\uparrow}-n_{j\downarrow}
\end{equation}
where the hard-core constraints have to be properly taken into
account. In this approximation, the BHM Hamiltonian is given by
\begin{eqnarray}
\label{EFM3}
H &=& -J\sum_j\left[2 c_{j\uparrow}^\dagger c_{j+1\uparrow}+c_{j\downarrow}^\dagger c_{j+1\downarrow} +h.c.\right] \nonumber \\
&+& \sqrt{2}J\sum_j\left[c_{j\uparrow}^\dagger c_{j+1\downarrow}^\dagger + c_{j\downarrow}^\dagger c_{j+1\uparrow}^\dagger + h.c.\right] \nonumber \\
&-& \frac{U}{2}\sum_j (n_{j\uparrow}+n_{j\downarrow}) + V\sum_j n_{j\uparrow}n_{j\downarrow}
\end{eqnarray}
with $V\to\infty$ required to project out unphysical states where a
holon and a doublon occupy the same site. In the following we drop
this constraint which is a reasonable lowest order approximation if
the number of holons and doublons in the system is very small. To
derive the EOM's, we can either diagonalize the Hamiltonian first by a
Fourier and a Bogoliubov transform or work directly with the
Hamiltonian \eqref{EFM3} in position space. We choose to do the latter
here, in which case we also have to consider the EOM's for the 'pairing
terms', see second line of \eqref{EFM3}. We introduce the following
shorthand notation: $h_{kl}=\langle c_{k\downarrow}^\dagger
c_{l\downarrow}\rangle$, $d_{kl} = \langle c_{k\uparrow}^\dagger
c_{l\uparrow}\rangle$, $a_{kl} = \langle c_{k\downarrow}
c_{l\uparrow}\rangle$, and $\bar a_{kl} = \langle
c_{k\downarrow}^\dagger
c_{l\uparrow}^\dagger\rangle=-a_{kl}^\dagger$. For the doublon
correlator the EOM then reads
\begin{eqnarray}
\label{EFM4}
\im\dot d_{kl} &=& 2J(d_{k-1l}-d_{kl+1}+d_{k+1l}-d_{kl-1}) \nonumber \\
&-& \sqrt{2} J (a_{k+1l}+\bar a_{l+1k} - \bar a_{l-1k}- a_{k-1l}) \nonumber \\
&-& \im\gamma_0 (\delta_{k0}+\delta_{l0})d_{kl} (1-\langle n_{0\downarrow}\rangle)
\end{eqnarray}
and for the holon
\begin{eqnarray}
\label{EFM5}
&&\im\dot h_{kl} = J(h_{k-1l}-h_{kl+1}+h_{k+1l}-h_{kl-1}) \nonumber \\
&-& \sqrt{2} J (-a_{lk+1}-\bar a_{kl+1} + \bar a_{kl-1}+ a_{lk-1})
\nonumber \\ 
&+& \im\gamma_0 [\delta_{k0}\delta_{l0}-\frac{1}{2}(\delta_{k0}+\delta_{l0})h_{kl}](1-\langle
n_{0\uparrow}\rangle) \nonumber \\ 
&-& \im\gamma_0\sqrt{2}\delta_{l0}(1-\delta_{k0})\bar
a_{kl}(1-\langle n_{0\uparrow}\rangle)(1-\langle
n_{0\downarrow}\rangle) \nonumber \\
&+& \im\gamma_0\sqrt{2}\delta_{k0}a_{lk}(1-\langle
n_{0\uparrow}\rangle)(1-\langle n_{0\downarrow}\rangle).
\end{eqnarray}
Note that the Hubbard interaction in this approximation is just a
chemical potential for the holons and doublons, see Eq.~\eqref{EFM3},
and therefore does not show up in the EOM's for these particles. The
Hubbard interaction does, however, show up in the EOM's for the
non-particle conserving, anomalous correlators which are given by
\begin{eqnarray}
\label{EFM6}
&& \im \dot a_{kl} = -J(a_{k+1l} + a_{k-1l} + 2 a_{kl+1}+ 2 a_{kl-1}) \nonumber \\
&-& \sqrt{2} J (d_{k-1l}+h_{l+1k} -\delta_{kl+1} - h_{l-1k} + \delta_{kl-1} - d_{k+1l}) \nonumber \\
&-& U a_{kl} -\im\gamma_0 \sqrt{2}\delta_{k0} d_{kl}(1-\langle n_{0\downarrow}\rangle)(1-\langle n_{0\uparrow}\rangle) \nonumber \\
&-&\im\gamma_0/2[\delta_{k0}a_{kl}(1-\langle n_{0\uparrow}\rangle)+ 3\delta_{l0}a_{kl}(1-\langle n_{0\downarrow}\rangle)]. 
\end{eqnarray}
The system of EOM's, Eqs.~(\ref{EFM4}-\ref{EFM6}), can then be solved
by numerical integration. Note that this approach is also a mean-field
decoupling scheme---although different from the one discussed in
Sec.~\ref{Decoupling}---based on restricting the local Hilbert space
to three states only and ignoring the infinite repulsion $V$ between
holons and doublons in Eq.~\eqref{EFM3} which is required to avoid
unphysical states with a holon and a doublon occupying the same
site. In the EOM's (\ref{EFM4}-\ref{EFM6}) this constraint is only
implemented 'on average'.

\subsection{LCRG and quantum trajectories}
\label{Sec_LCRG}
The quantum trajectory (QT) approach was developed in the 1990's
\cite{Carmichael1993,Molmer1993} as a wave-function approach to
dissipative processes in quantum optics. The term quantum trajectories
was coined by M. Carmichael.\cite{Carmichael1993} Previously, it was
called either the quantum jump approach or the Monte Carlo
wave-function method. In general the QT approach can be used to solve
any master equation in Lindblad form.\cite{Daley2014} An integration
over these QT's can be carried out by any numerical approach that is
able to solve the Schr\"odinger equation.

The main idea is to rewrite the master equation as a stochastic
average of QT's. Each QT is dependent on random variables, thus no QT
is identical in the limit time $t$ to infinity. The main advantage of
the QT approach is that only a quantum state has to be evolved in time
thus avoiding to propagate the full density operator; only a Hilbert
space with the dimension of the system needs to be
considered.\cite{Carmichael1993,Daley2014,Molmer1993} The prize one
has to pay is that many QT's (of the order of several hundred or even
several thousand for the Bose-Hubbard model) have to be calculated to
obtain meaningful averages. While memory requirements are reduced and
obtainable simulation times often greatly enhanced as compared to a
direct time evolution of the density matrix, the QT approach is
therefore very costly in terms of computing time.

\begin{figure}[!ht]
\includegraphics*[width=0.98\linewidth]{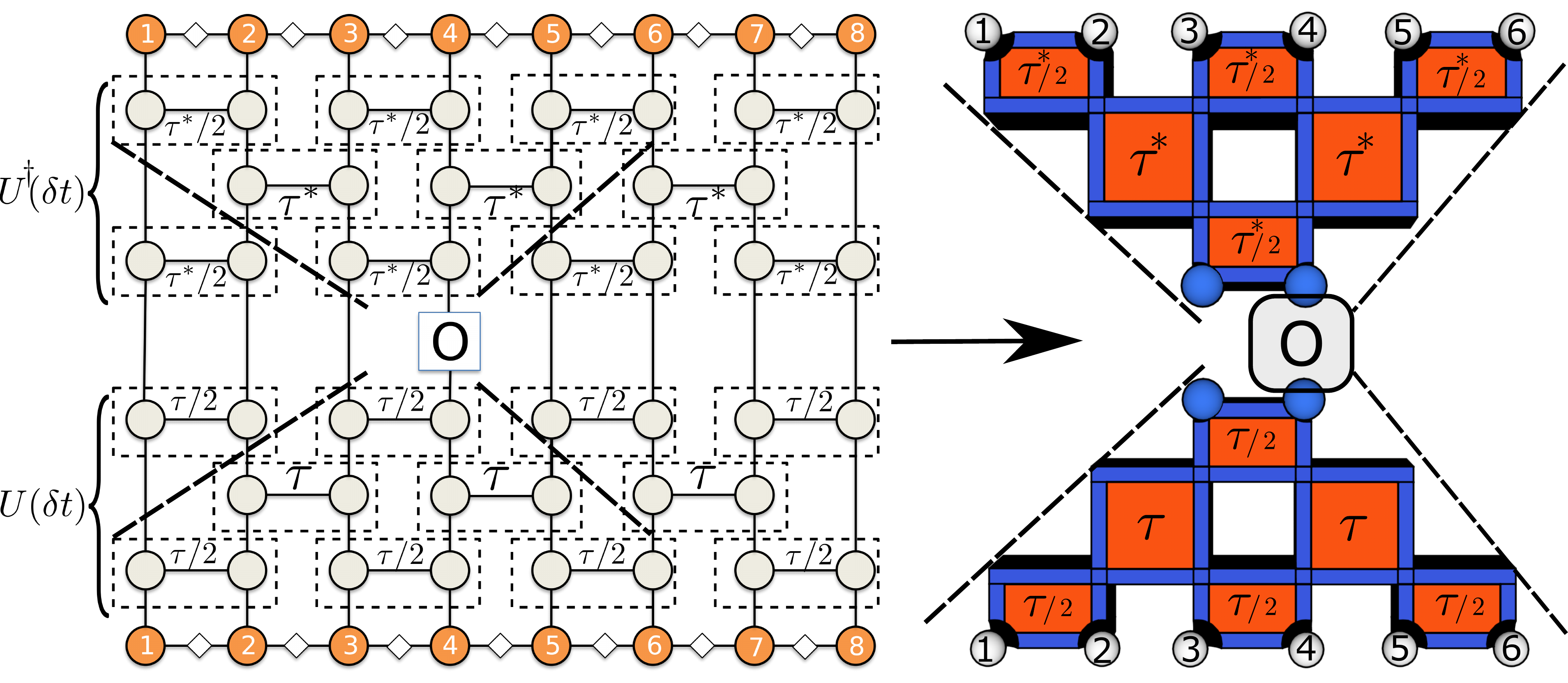}
\caption{Transformation from an MPS network after Trotter-Suzuki decomposition (left) 
to a light cone (right). The adjacent plaquettes outside the light
cone, depicted by the black dotted line, cancel each other.}
\label{Fig_MPS-Lightcone}
\end{figure}

In previous studies of dissipative one-dimensional quantum systems,
the QT approach has been combined with time-dependent DMRG algorithms
for finite system size $L$ \cite{Barmettler2011} and compared to the
superoperator approach where the full density matrix is evolved in
time.\cite{BonnesCharrier,BonnesLauchli} One problem with numerical
algorithms for finite systems is that the only non-equilibrium steady
state (NESS) which can be reached in a system with particle loss is
the vacuum. We circumvent this problem here by combining the QT
approach with the LCRG making it possible to discuss the physical
properties at long times $t\gg J/\gamma$. The LCRG uses a
Trotter-Suzuki decomposition of the time evolution operator\cite{Suzuki1976} and the
Lieb-Robinson bounds\cite{LiebRobinson,BravyiHastings} to restrict the
time-evolution to an effective light cone for a Hamiltonian with short
range interactions.\cite{EnssSirker} The LCRG algorithm efficiently
simulates one-dimensional systems and yields observables directly in
the thermodynamic
limit.\cite{AndraschkoEnssSirker,AndraschkoSirker2,EnssAndraschkoSirker}
The Hilbert space is truncated based on the eigenvalues of the reduced
density matrix following traditional DMRG
schemes.\cite{White1992,WhiteFeiguin} Transfer matrices are used to
enlarge and time-evolve the system stepwise. In an alternative
description, a tensor network in matrix product state (MPS) language
can be easily transformed to a light cone shape through this process,
see Fig.~\ref{Fig_MPS-Lightcone}.

The aim of this section is to describe how to combine the LCRG with a
QT scheme in order to generalize the LCRG algorithm to non-unitary
time evolution in open systems for the special case where a single
site is coupled to a bath. The LCRG keeps its light-cone shape for a
non-unitary time-evolution carried out only locally. Translational
invariance, however, is destroyed thus algorithms like the infinite
TEBD
\cite{VidaliTEBD,Orus2008} cannot directly be applied to calculate the dynamics
in the thermodynamic limit. 

The size of the effective light cone in the LCRG algorithm at time $t$
is given by
\begin{equation}
L = v_{\textrm{Trotter}}|t|,\quad v_{\textrm{Trotter}}=\frac{a}{\delta t},
\end{equation}
where $v_{\textrm{Trotter}}$ is the Trotter speed and $a$ the lattice
constant which we have set to unity. It has been shown that for
systems with short-range interactions the error between a
time-dependent operator $\hat{O}_{0}^{(\ell)}(t)$ acting on the site
$j=0$ in the middle of an $\ell$ site light cone (see
Fig.~\ref{Fig_MPS-Lightcone} and the operator $\hat{O}_{0}(t)$
evaluated in the infinite system is bounded by
\begin{equation} 
\| [\hat{O}_0(t),\hat{O}_{0}^{(\ell)}(t)]  \| \leq c e^{\left(-\frac{\ell-v_{\textrm{LR}}t}{\xi}\right) },
\label{barvyibound}
\end{equation}
where $c$ is a constant, $\xi$ the correlation length and
$v_{\textrm{LR}}$ the Lieb-Robinson velocity which is of order of the
hopping amplitude $J$ and describes the velocity information is
spreading through the lattice.\cite{LiebRobinson,BravyiHastings} In
the BHM the Lieb-Robinson bound has been observed in tDMRG
simulations\cite{BarmettlerPoletti} and has also been verified in
experiment.\cite{CheneauBarmettler} In order to make the error in the
LCRG simulations exponentially small as compared to results in the
thermodynamic limit we therefore have to make sure that the condition
$v_{\textrm{Trotter}}\gg v_{\textrm{LR}}$ is fulfilled. For the
one-dimensional BHM in the limit of $U/J\rightarrow
\infty $, doublon and holon excitations with velocities 
$v_{\mathrm{doublon}}= 4 J$ and $v_{\mathrm{holon}}= 2 J$
respectively exist.
\cite{CheneauBarmettler,AndraschkoSirker2} For a propagating doublon 
the Trotter time step therefore has to be chosen such that
\begin{equation}
1/\delta t \gg v_{\mathrm{doublon}}\sim 4 J.
\label{trotterlieb}
\end{equation} 
In our simulations we usually set $J\delta t \sim 0.01$ or smaller,
which obviously fulfills Eq.~(\ref{trotterlieb}). We thus obtain
results in the thermodynamic limit with the light-cone structure only
introducing exponentially small errors.

The QT approach can then be combined with the LCRG algorithm in the
following way: The system without dissipation evolves under a
Hamiltonian $H=\sum_j h_{j,j+1}$. For a system with hopping terms or
interactions beyond nearest-neighbors the unit cell has to be expanded
accordingly. Adding local dissipation at site $k$ we have to replace
the local Hamiltonian by $h_{j,j+1}\to
h^{\textrm{eff}}_{j,j+1}=h_{j,j+1}-\delta_{jk} \frac{\gamma_k}{2} \hat
A_k^\dagger \hat A_k$. The local time evolution operator in
Trotter-Suzuki decomposition is then given by $\tau =\exp(-\im\delta t
h^{\textrm{eff}}_{j,j+1})$ and is depicted as a plaquette in
Fig.~\ref{Fig_MPS-Lightcone}. Next, we draw a random number $r\in
[0,1)$. The normalized initial state is now time-evolved $|\Psi(\delta
t)\rangle =
\exp(-\im H^{\textrm{eff}}\delta t)|\Psi(0)\rangle $. If $r<\|
|\Psi(\delta t)\rangle \| $ we continue with the time evolution. If,
on the other hand, $r\geq \| |\Psi(\delta t)\rangle \| $ then we apply
the local operator $A_k$ onto the state, $|\Psi(\delta t)\rangle \to
A_k|\Psi(\delta t)\rangle $, realizing a quantum jump. After the
quantum jump the time evolved state is normalized, a new random
variable $r\in [0,1)$ is drawn, and the state is further evolved in
time under the effective Hamiltonian until the next quantum jump
occurs. In the implementation it is important to use very small time
steps close the point where the quantum jump occurs in order to avoid
having many trajectories which jump at exactly the same
time.\cite{Barmettler2011,BonnesLauchli} For each QT the expectation
value of the variable of interest is measured and then averaged over
all QT's. The statistical error of an observable is simply given by
$\sigma_A(\hat{O}) =
\frac{\sigma(\hat{O})}{\sqrt{Q}} $, where $Q$ is the number of QT and
$\sigma(\hat{O})$ the standard deviation because the QT's are
statistically independent.\cite{Daley2014,Molmer1993} For the density
and current profiles we typically need several hundred QT's to obtain
statistical errors which are small compared to the dynamical changes
of the observables. The number of states $\chi$ which we need to keep
is adjusted dynamically so that the truncation error always stays
below $10^{-7}$. This typically requires the number of states to be in
the range $\chi\sim 300-1850$ for the examples considered
later. Furthermore, we always recompute each QT with a higher bond
dimension to make sure that all QT's are converged as one fixed $\chi$
does not apply to all QT's.

\subsection{Preparation of initial state and comparison with exact diagonalization}
We consider dissipative dynamics starting from the ground state of the
closed system. In order to compute the ground state within the LCRG
scheme, imaginary time evolution is used. Because the Bose-Hubbard
model lacks particle-hole symmetry, the correct chemical potential
$\mu$ needs to be included in the imaginary time evolution operator
\begin{equation}
\tau_\beta = \exp\big\{-\beta(H- \mu \sum\limits_j n_j)\big\}.
\end{equation}
Here $\beta$ is imaginary time. The time evolution operator is then
considered in Trotter-Suzuki composition and $\beta$ successively
increased to project an arbitrary initial state onto the ground
state. In practice, we cannot reach the limit $\beta\to\infty$ so that
the projection will not be exact. However, if there is a gap
$\Delta_E$ between the ground state and the first excited state then
the error will be exponentially small, $\sim \exp(-\beta\Delta_E)$, if
$\beta\Delta_E\gg 1$.

\begin{figure}[!ht]
\center
\includegraphics*[width=0.99\columnwidth]{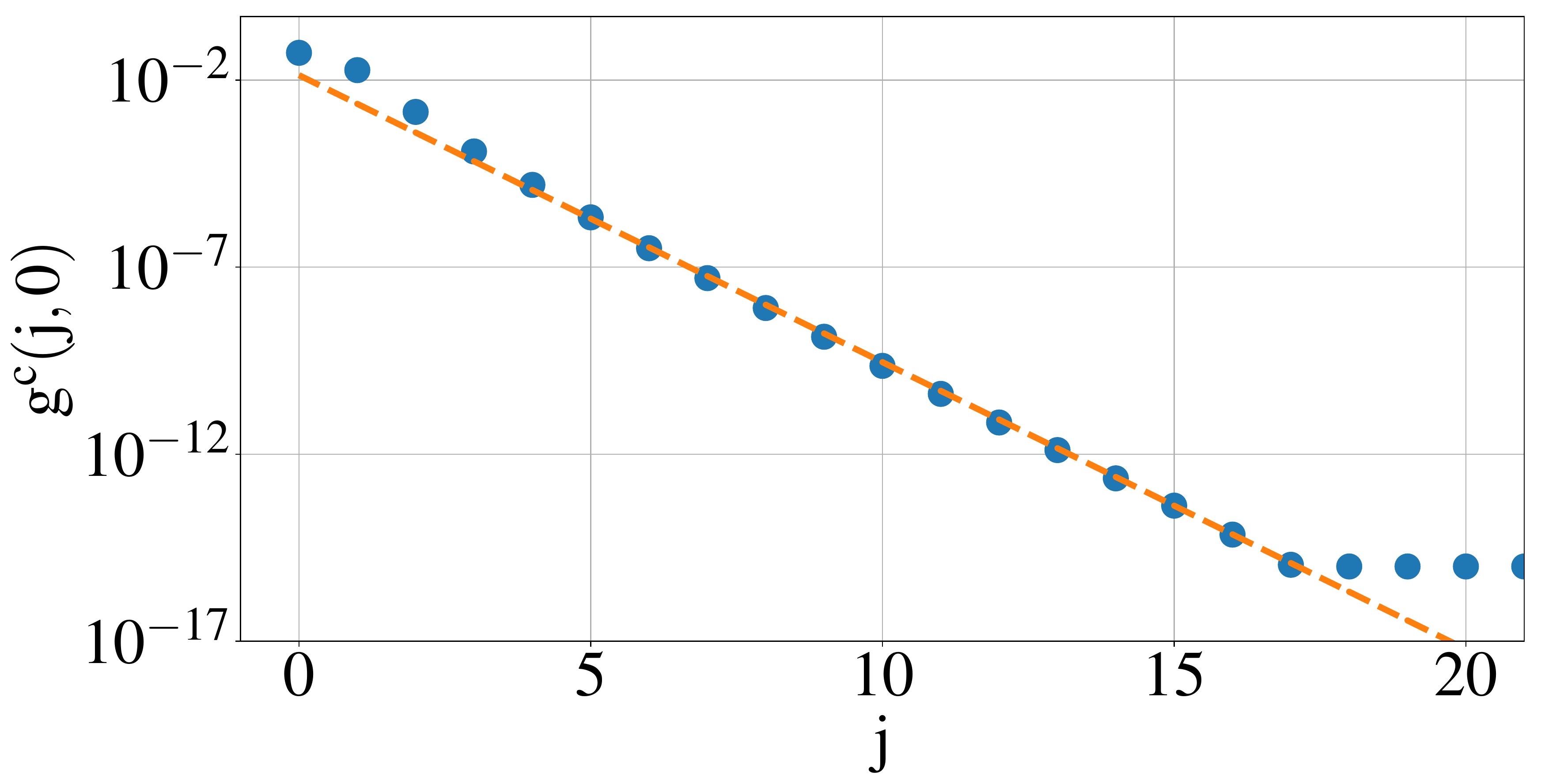}
\caption{Connected density-density correlation function, Eq.~\eqref{gc}, 
in the projected state for $U=12J$: $g^c(j,0)$ is decaying
exponentially with a correlation length $\xi\approx 0.57$. The line is
an exponential fit. Correlations for $j>17$ are of the order or
smaller than $10^{-16}$ and are therefore not correctly reproduced in
double precision.}
\label{Fig_GS_corr}
\end{figure}

As an example, we consider imaginary time evolution deep in the
Mott-insulating phase for $U=12J$. For $\beta\sim 5$, we find that the
energy of the state is already converged. To further test the
properties of the projected state, we consider the connected
density-density correlation function
\begin{equation}
\label{gc}
g^c(j,t)=\langle n_0 n_j\rangle(t) - \langle n_0\rangle\langle n_j\rangle(t).
\end{equation}
As shown in Fig.~\ref{Fig_GS_corr}, the projected state is not a
simple product state but rather shows the physically expected
exponentially decaying correlations.
We ensure that the energy is converged for all the projected ground
states considered in the following and that the correlations are
properly captured.

\begin{figure}[!ht]
\includegraphics*[width=0.98\linewidth]{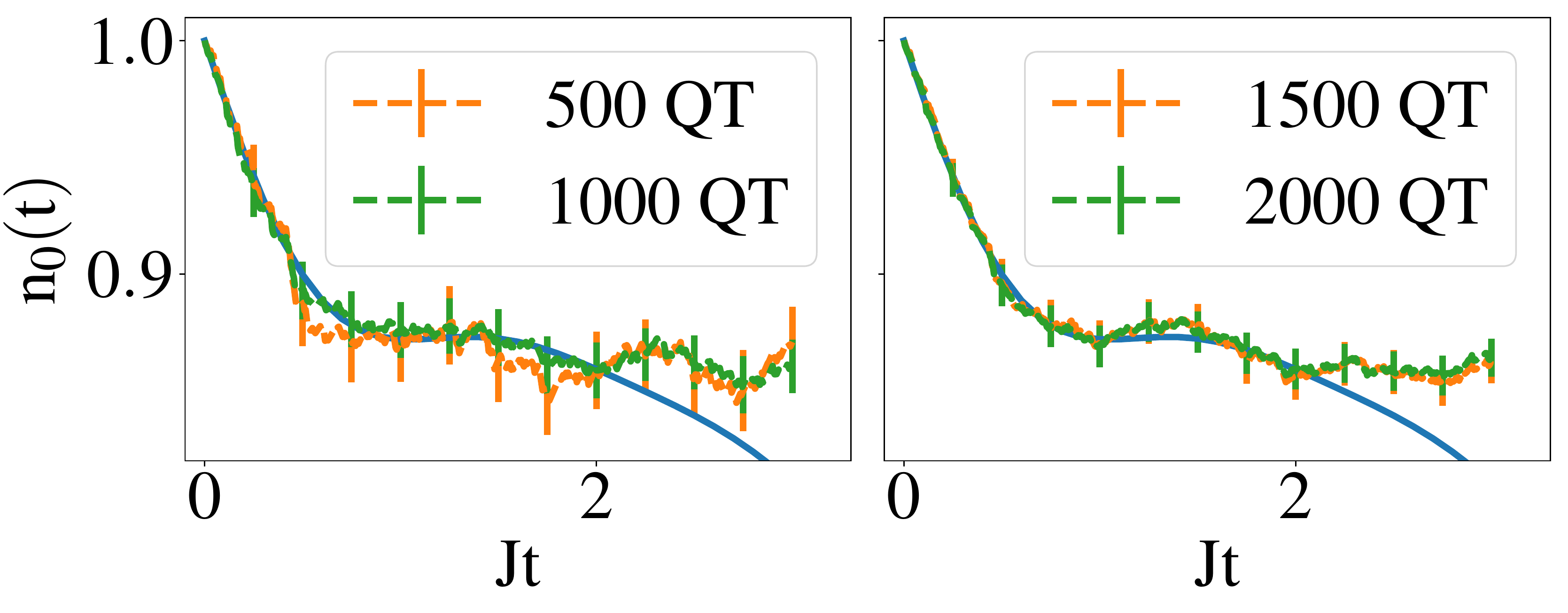}
\caption{Time evolution of the density $n_0(t)$ at the dissipative site starting from the 
Mott-insulating state with $U=12J$. Compared are QT-LCRG results
(error bars denote the statistical errors) with exact diagonalization
data (solid line) for a chain of $L=5$ sites and
$\gamma=0.25$. Boundary effects for $n_0(t)$ become visible in the ED
results for $Jt\geq 2$.}
\label{Fig_Comparison_ED}
\end{figure}

As a next step, we test the QT-LCRG algorithm by comparing results for
the BHM with local particle loss with a solution of the Lindblad
equation \eqref{LME2} based on exact diagonalizations (ED). Note that
such a comparison is only meaningful for observables at or very close
to the lossy site (which will always be in the middle of the
considered chain) and small times because ED is limited to very small
system sizes. In Fig.~\ref{Fig_Comparison_ED} results for the density
$n_0(t)$ at the lossy site calculated with the QT-LCRG algorithm using
$500-2000$ QT's are compared to the ED result.
Within the statistical errors, both results agree for $Jt\leq 2$.

\section{Results}
\label{Results}
In the following, we want to analyze results obtained by the QT-LCRG
algorithm for the density and current profiles, the density-density
correlations, as well as the entanglement entropy. We will compare
these results to the EOM approach and are, in particular, interested
in the long-time regime where many-body effects dominate and the EOM
approach in Hartree-Fock approximation is expected to fail. We will
mainly concentrate on the case of weak dissipation but will also
briefly discuss the case of strong dissipation towards the end of this
section.

\subsection{Particle and density profiles}
\begin{figure}[!ht]
\includegraphics*[width=0.99\columnwidth]{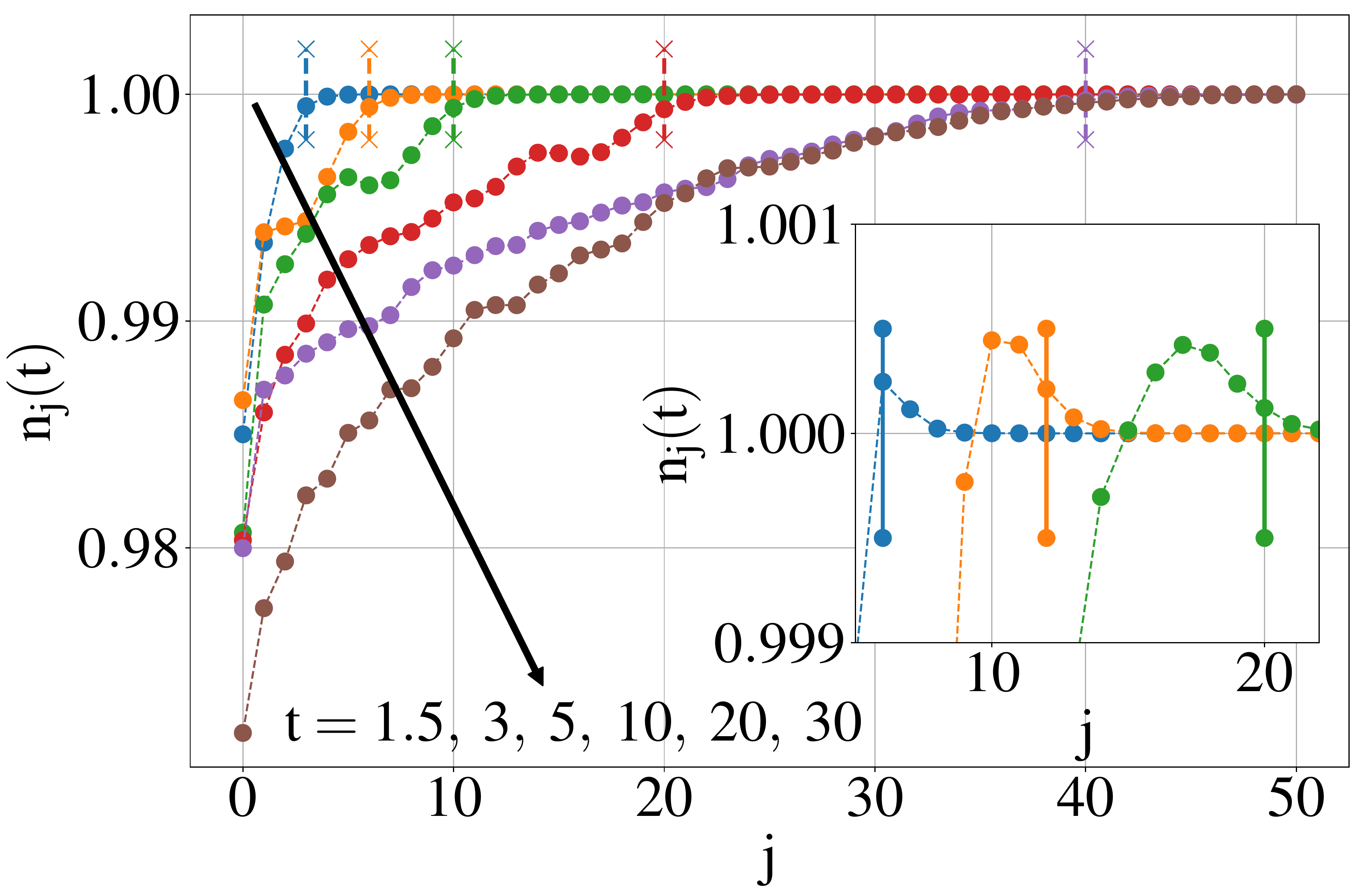}
\caption{Density profiles at short and intermediate times for $U=12$ and $\gamma = 0.025$. 
The holons spread in a light-cone like fashion with the numerically
calculated velocity $v\approx 2J$ (vertical
bars).\cite{AndraschkoSirker2} Averages over $2589$ converged QT's for
$t\leq 20$ and $1089$ for $t> 20$ are shown for bond dimensions
$\chi=900-1450$. The statistical error is largest at the dissipative
site, $\sigma^0_A \approx 0.0028$. Inset: For $U=12$, $\gamma=8$ a
doublon contribution is clearly visible with $v\approx 4J$ (vertical
bars).}
\label{Fig_density_short_times}
\end{figure}

When a hole is created at the lossy site $j=0$, this density
perturbation starts to move through the lattice with the holon
velocity which is approximately given by $v\sim 2J$ for $U/J\gg
1$. Based on the effective fermion model description,
Eq.~\eqref{EFM3}, we see that alternatively also a doublon can be
annihilated---although the doublon density in the initial Mott
insulating state with $\langle n_j\rangle =1$ will be small---,
creating a perturbation which will travel with twice the holon
velocity.\cite{AndraschkoSirker2} For small dissipation and large $U$
we cannot reliably detect the doublon contribution numerically so that
the density profile has a light cone structure at short times given by
the holon velocity, see Fig.~\ref{Fig_density_short_times}. However,
the doublon contribution is present and can be detected numerically
for larger $\gamma$ values, see the inset of
Fig.~\ref{Fig_density_short_times}.

The changes of the density profiles are caused by local currents which
can be calculated from the continuity equation $\langle\dot n_j\rangle
=-\langle\textrm{div}\, \mathcal{J}\rangle = -(\langle
\mathcal{J}_j\rangle -\langle
\mathcal{J}_{j-1}\rangle)$ with
\begin{eqnarray}
\label{ndot}
\dot n_j = \dot\sigma_{jj} &=& \im J(\sigma_{j,j+1}-\sigma_{j+1,j}+\sigma_{j,j-1}-\sigma_{j-1,j}) \nonumber \\
&-& \gamma_0 \delta_{j0}\sigma_{jj},
\end{eqnarray}
see Eq.~\eqref{EOM2}. The local current operator originating from the
unitary part is therefore given by
\begin{equation}
\label{current}
\mathcal{J}_j = -\im J\left(b^\dagger_j b_{j+1} - b^\dagger_{j+1}b_j\right).
\end{equation}
Using the current operator, the change of the local density can also be
written as
\begin{equation}
\label{ndot2}
\frac{d}{dt}\langle n_j \rangle = \left\{\begin{array}{cc} \langle \mathcal{J}_{j-1}\rangle - \langle \mathcal{J}_j\rangle -\gamma_0 \langle n_0\rangle & \quad i=0 \\*[2.5pt] \langle \mathcal{J}_{j-1}\rangle - \langle \mathcal{J}_j\rangle & \quad\text{else} \end{array} \right. 
\end{equation}
with $\langle \mathcal{J}_j\rangle = 2\,\text{Im}\,\langle b^\dagger_j
b_{j+1}\rangle$. At short times inside the light cone we expect that
$\frac{d}{dt}\langle n_j\rangle <0$ which is equivalent to $\langle
\mathcal{J}_j\rangle > \langle \mathcal{J}_{j-1}\rangle$ for $|j|>1$. For $j>1$ 
(to the right of the dissipative site) at the boundary of the
light-cone we expect $0\approx |\langle
\mathcal{J}_j\rangle| < |\langle \mathcal{J}_{j-1}|\rangle $ which implies that the
currents are negative, i.e., are pointing towards the dissipative
site. As long as the local densities inside the light cone are
decreasing we furthermore expect that the local currents $\langle
\mathcal{J}_{j>0}\rangle$ are a monotonically increasing function of the
distance $j$ from the dissipative site. Because the Hamiltonian is
reflection symmetric around $j=0$ there is always a current equal and
opposite in direction on the other side of the lossy site
\begin{equation}
\label{current2}
\langle \mathcal{J}_j\rangle(t) = -\langle \mathcal{J}_{-j-1}\rangle (t).
\end{equation}
Furthermore, we can also immediately read off the stationary current
from Eq.~\eqref{ndot2} by demanding $\frac{d}{dt}\langle n_j\rangle
=0$ for all sites $j$. Using Eq.~\eqref{current2} this leads to
\begin{equation}
\label{current3}
\langle \mathcal{J}_j \rangle_{\textrm{NESS}} = \left\{\begin{array}{cc} -\gamma_0\langle n_0\rangle/2 & \quad j\geq 0 \\ \gamma_0\langle n_0\rangle/2 & \quad j<0 \end{array} \right. \, .
\end{equation}

\begin{figure}[!ht]
\includegraphics*[width=0.99\columnwidth]{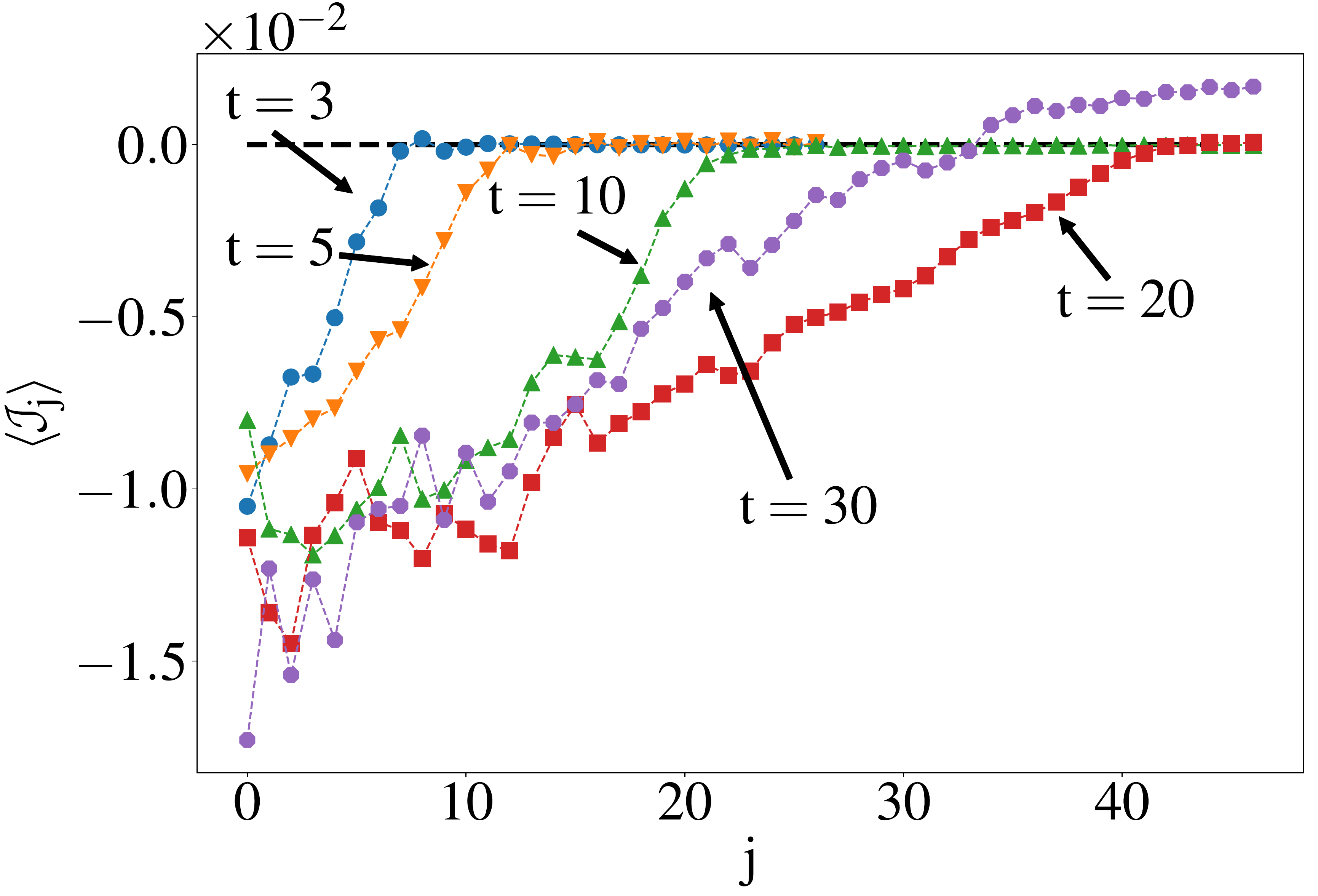}
\caption{Current profiles at short and intermediate times for $U=12$ and 
$\gamma = 0.025$ with $1500$ converged QT's for $t\leq 20$ and $574$
for $t> 20$. At times $t>20$ a current reversal at sites far away from
the lossy site is starting.}
\label{Fig_currents_short_times}
\end{figure}

At short times our numerical results for the currents are consistent
with these considerations, see Fig.~\ref{Fig_currents_short_times}. 
We have also checked that the density and current profiles are
consistent with the continuity equation
\eqref{ndot2}. For intermediate times we find that the area of reduced density first
continues to spread before essentially stopping to extend further at
times $t\sim 30$. As shown in Fig.~\ref{Fig_density_short_times} the
density at this point is significantly reduced on the first $\sim 40$
lattice sites around the defect. The current profiles shown in
Fig.~\ref{Fig_currents_short_times} also show an intricate evolution
at this timescale. For the $\sim 10$ sites closest to the defect the
currents, on average, stop growing for $t\geq 10$. Even more
remarkable, a local current reversal starts to set in for times $t>
20$ at sites further away from the defect. For time $t=30$, for
example, sites $j>35$ have a local current {\it leading away} from the
lossy site. This current reversal will ultimately reduce the area over
which the density is significantly depleted while even further
reducing the density close to the defect.

\begin{figure}[!ht]
\includegraphics*[width=0.99\columnwidth]{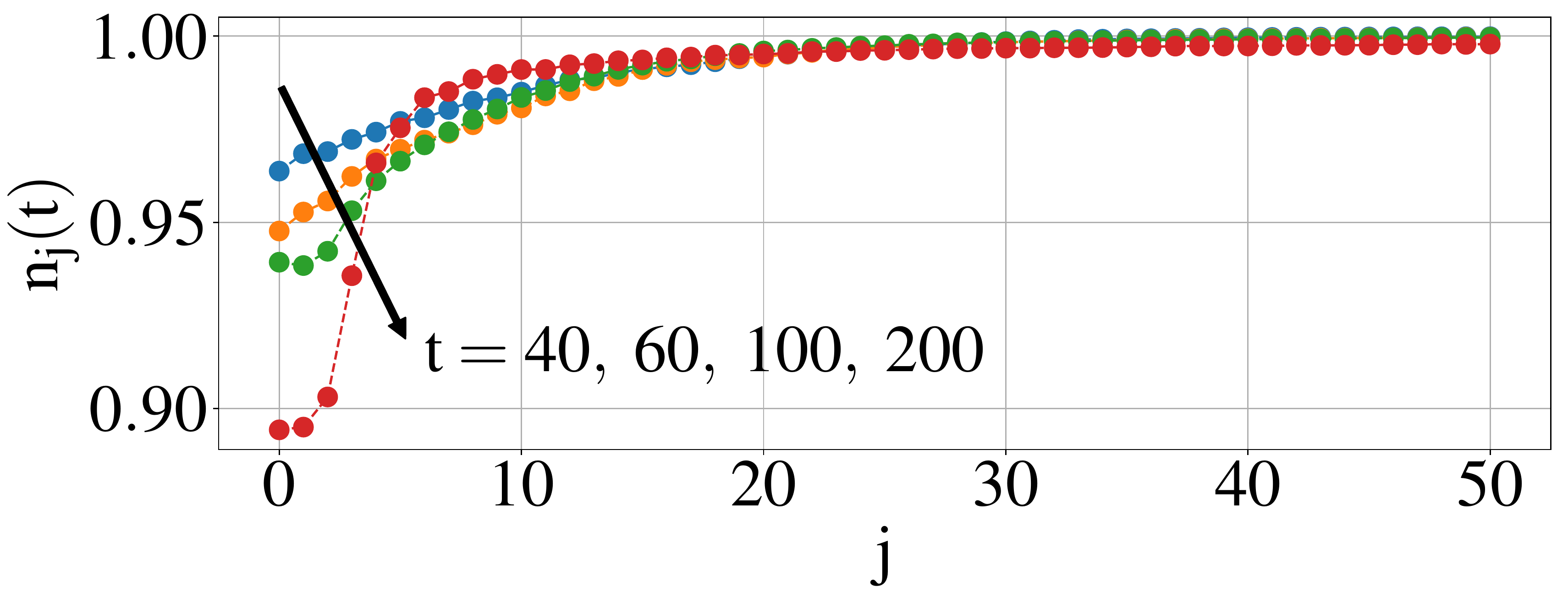}
\caption{Density profiles at long times for $U=12$ and $\gamma = 0.025$. The density profile at sites 
$j\gtrsim 10$ changes only very little over time.}
\label{Fig_density_long_times}
\end{figure}

\begin{figure}[!ht]
\includegraphics*[width=0.99\columnwidth]{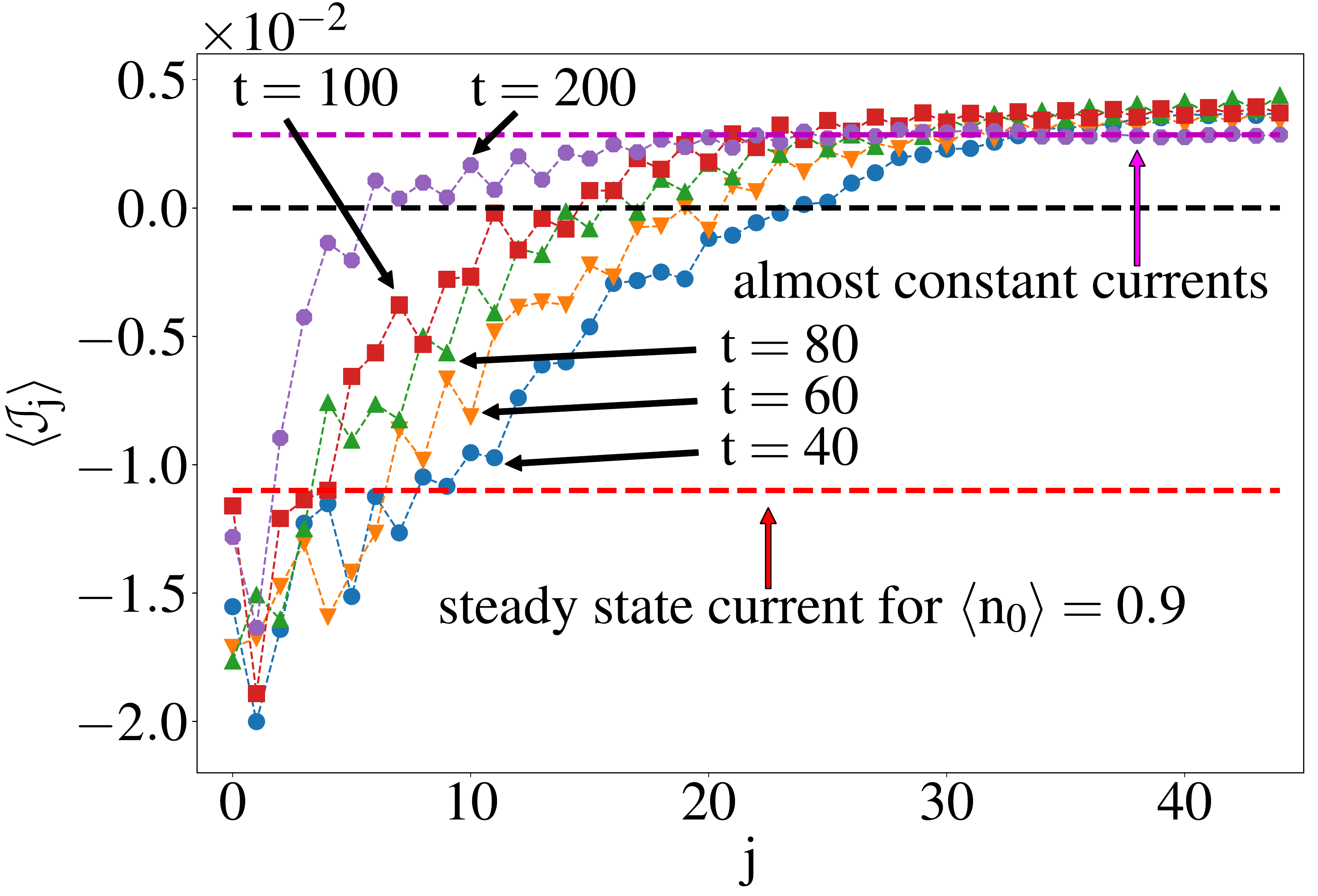}
\caption{Current profiles at long times for $U=12$ and $\gamma = 0.025$. 
The local currents at $t=200$ are reversed and are pointing away from
the lossy site.}
\label{Fig_currents_long_times}
\end{figure}

\begin{figure}[!ht]
\includegraphics*[width=0.99\columnwidth]{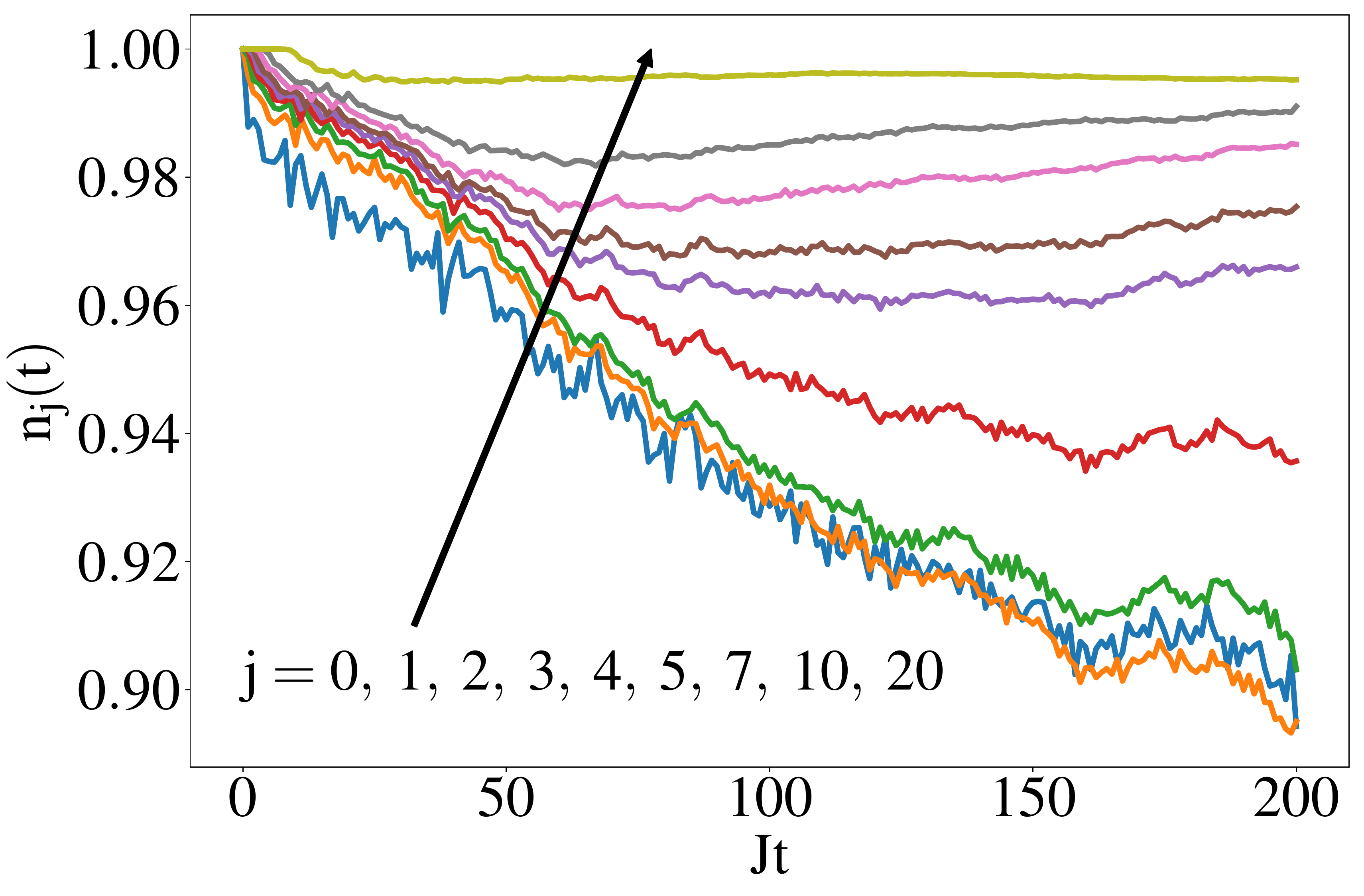}
\caption{Local densities as a function of time for weak local particle loss $\gamma = 0.025$. 
The local densities change very little in the time interval $t\in [150,200]$ pointing to a metastable steady state.}
\label{Fig_local_densities}
\end{figure}

This effect can clearly be seen in the density profiles at the longest
simulation times shown in Fig.~\ref{Fig_density_long_times}.
The densities at sites $j\gtrsim 10$ only change very little in time
while the densities at the sites closest to the defect continue to be
reduced. The absolute values of the currents near the defect also do
decrease at long times as shown in Fig.~\ref{Fig_currents_long_times}.
Interestingly, all local currents for $j>5$ are reversed at $t=200$
and are flowing away from the defect. While the local currents are
almost all equal as is expected in the steady state, see
Eq.~\eqref{current3}, the currents are flowing in the opposite
direction than the steady state currents. Furthermore, the magnitude
of the currents is much smaller than the steady state current for
$\langle n_0\rangle(t=200)\approx 0.9$, see
Fig.~\ref{Fig_currents_long_times}. We conclude that while the density
profile appears to become almost independent of time and the local
currents almost all equal to each other we are {\it not} in the
non-equilibrium steady state of the system. The local densities for
times $t\in [150,200]$, however, only change very little (see
Fig.~\ref{Fig_local_densities}) indicating that we have reached a
metastable steady state with small local currents pointing away from
the lossy site.
While the densities at sites $j=0-3$ at times $t\in [150,200]$ appear
to be constant on average, the particle densities at sites $j=5-10$
are monotonuously but very slowly increasing. Even further away from
the defect, on the other hand, the densities continue to decrease
slowly.

\begin{figure}[!ht]
\includegraphics*[width=0.99\columnwidth]{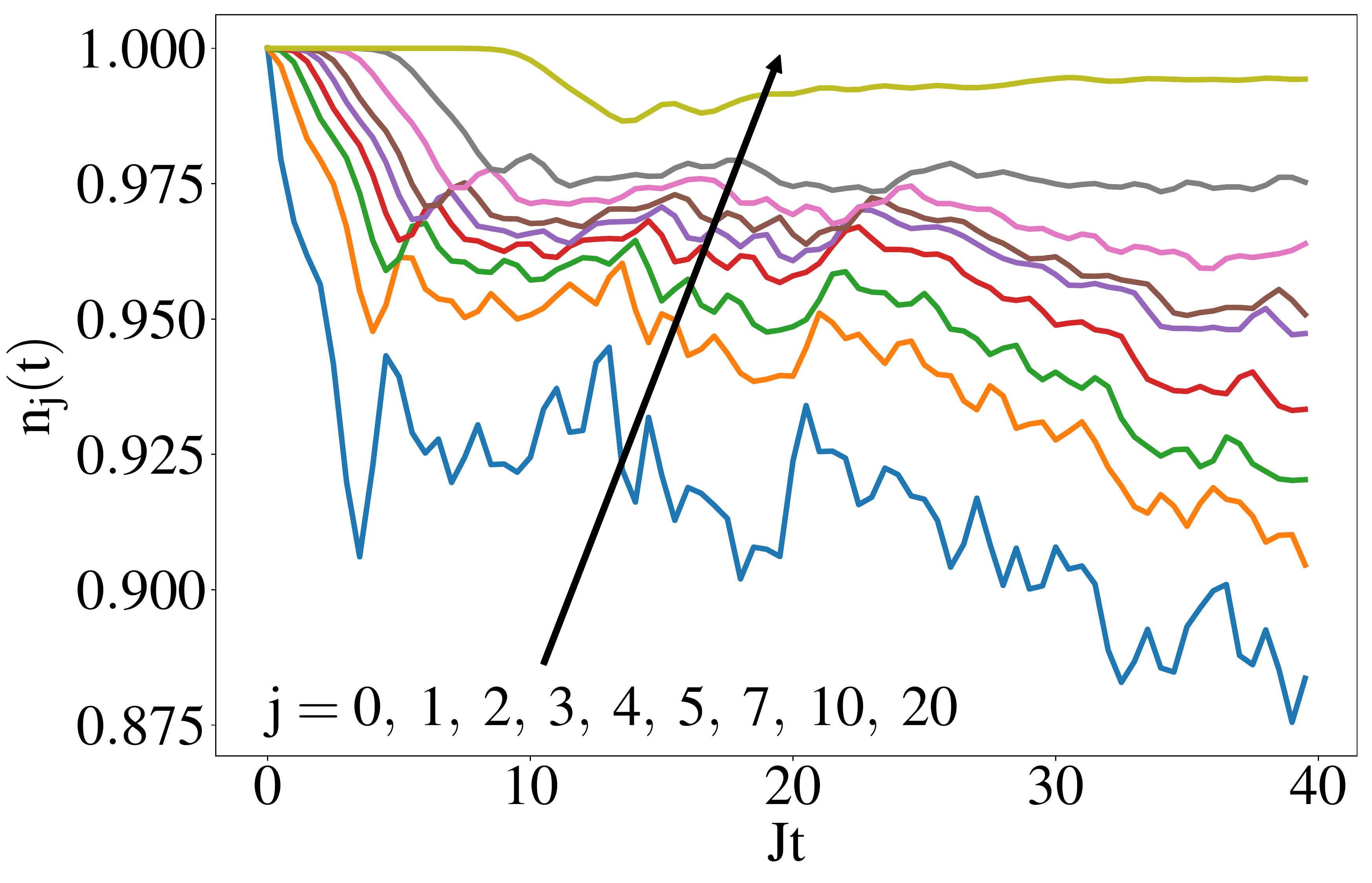}
\caption{Local densities as a function of time for $U=12$ and $\gamma = 0.1$. 
Averages over $800$ QT's are shown. At the time scale $J/\gamma=10$ a
plateau-like feature is visible.}
\label{Fig_densities_gamma0p1}
\end{figure}

While we cannot reach the non-equilibrium steady state, a likely
scenario based on the density and current profiles is a steady state
density profile which is quite steep with a density at the dissipative
site which is strongly reduced---and perhaps much closer to zero than
to $1$---while substantial particle densities persist on all other
sites. A small density at the dissipative site in the steady state
would, according to Eq.~\eqref{current3}, also lead to a small steady
state current. The local currents at times $t>200$ therefore possibly
stay almost equal except very close to the dissipative site but slowly
change sign again. Another surprising result of the simulations is the
very large time scale apparently required to reach the NESS. One
relevant time scale is clearly set by $J/\gamma = 40$ for the example
considered here. In Fig.~\ref{Fig_local_densities} this time scale
separates the regime where the densities of sites inside the light
cone change approximately linearly in time from a regime where the
densities at some sites become already approximately constant or even
start to slowly increase again. $J/\gamma$ is also the time scale
where some of the local currents start to reverse. In order to check
this interpretation we also briefly consider the case $\gamma=0.1$ in
the following. In this case the local densities $\langle
n_j\rangle(t)$ inside the light cone show an initial decay followed by
a plateau around $J/\gamma =10$ and then a further decay, see
Fig.~\ref{Fig_densities_gamma0p1}.
Again, the time scale to reach the NESS appears to be much larger than
$J/\gamma$.

\subsection{Comparison with EOM}
\begin{figure}[!ht]
\includegraphics*[width=0.99\columnwidth]{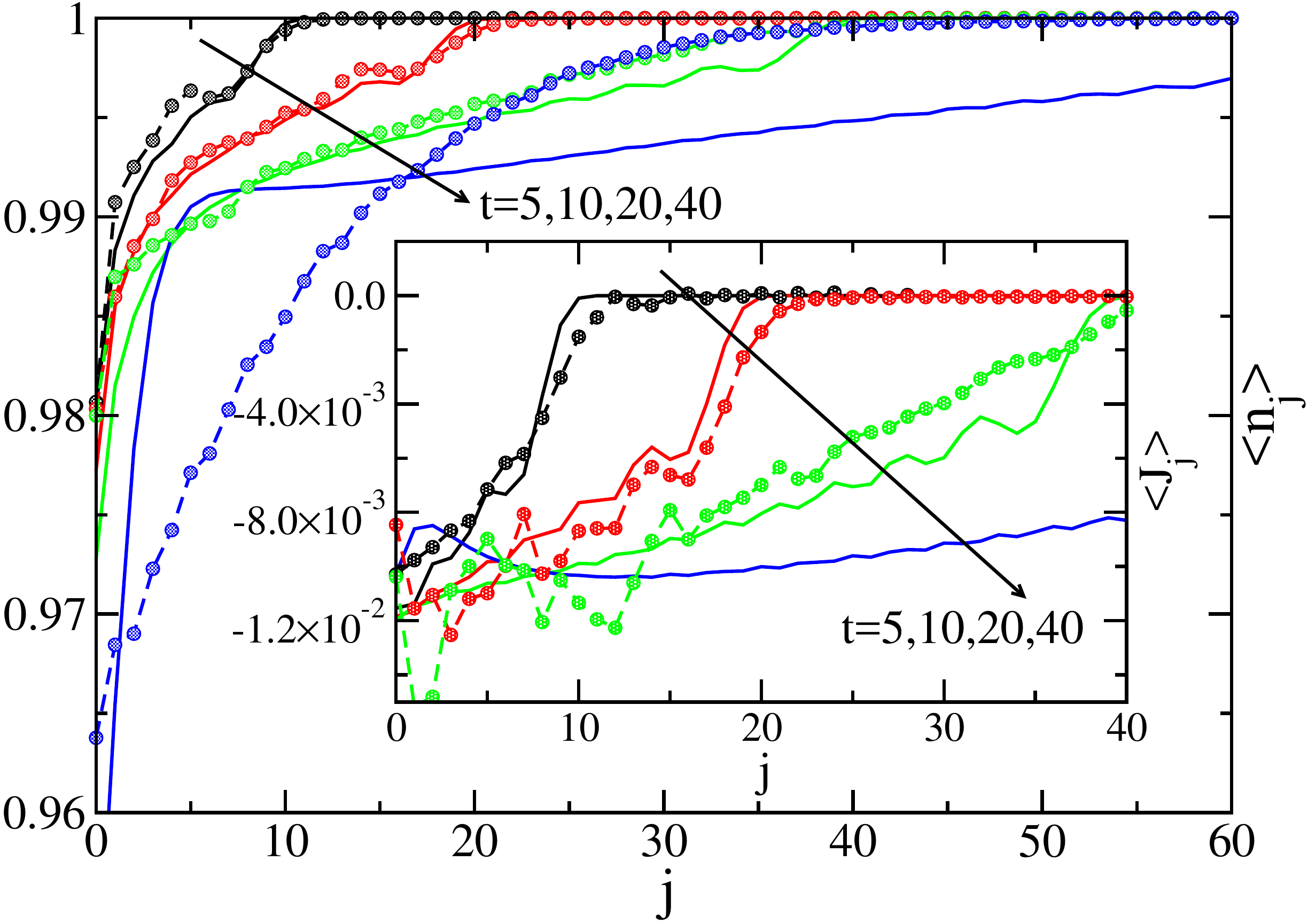}
\caption{Comparison of density profiles (main) and current profiles (inset) 
obtained by LCRG (symbols) and the EOM with Hartree-Fock decoupling
(lines), see Eq.~(\ref{EOM2}, \ref{EOM2p1}), for $U=12$ and $\gamma =
0.025$. LCRG current profiles are shown for $t=5,10,20$ only for
clarity.}
\label{Fig_EOM1}
\end{figure}

Next, we want to investigate how much of the complicated dynamics is
captured in an EOM approach with a Hartree-Fock decoupling or within
the effective fermion model approach, see Sec.~\ref{Sec_EOM}. In
Fig.~\ref{Fig_EOM1} the density and current profiles obtained by LCRG
and the first order Hartree-Fock EOM approach for short and
intermediate times are compared. As initial state in the EOM
calculations we use a product state with one boson per site which is a
good approximation for the ground state at $U=12$.
While the density profiles for times $t=5$ and $t=10$ obtained by this
mean-field EOM approach agree very well with the LCRG results, first
significant deviations become visible at $t=20$ and already at $t=40$
the Hartree-Fock EOM approach fails completely. While the LCRG data
show that between time $t=20$ and time $t=40$ the densities at sites
$j\geq 20$ no longer decrease, the EOM predicts a ballistic
extension of the region with reduced particle density with the holon
velocity $v\sim 2J$. That the EOM fails to capture essential aspects
of the dynamics is also obvious from the current profiles shown in the
inset of Fig.~\ref{Fig_EOM1}. While the current profiles are correctly
captured for $t<20$, the EOM approach completely fails for longer
times (see also Fig.~\ref{Fig_currents_long_times}). In particular,
current reversals away from the lossy site do not occur in the
mean-field EOM solution. We conclude that the current reversals
observed in the LCRG simulations are a genuine many-body effect which
cannot be captured in a Hartree-Fock decoupling scheme. The
Hartree-Fock solution---which is essentially the result for a Gaussian
system---is only able to describe the initial dynamics at times $t<
J/\gamma$.

\begin{figure}[!ht]
\includegraphics*[width=0.99\columnwidth]{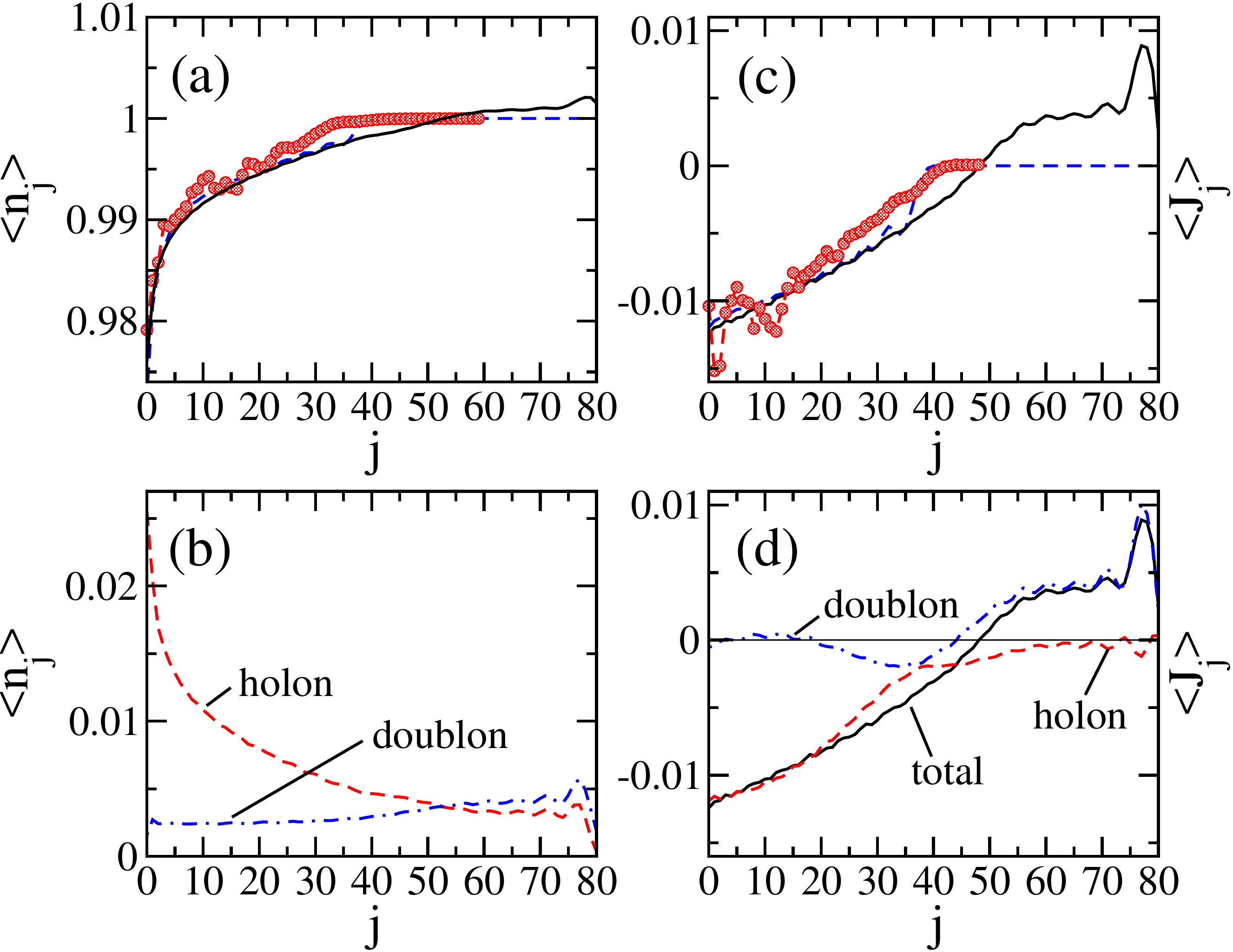}
\caption{(a) Density profiles from QT-LCRG (symbols), Hartree-Fock EOM (dashed line) and effective fermion model (solid line) at $t=20$. (b) Density of holons and doublons at $t=20$ in the EFM approach. (c,d) Same as (a,b) for the local currents.}
\label{Fig_HF_vs_EFM}
\end{figure}

The failure of the first order Hartree-Fock EOM decoupling scheme at
long times raises the question if a different EOM scheme can better
describe the system. We checked first that going to a second order
scheme as described by Eqs.~(\ref{EOM3}, \ref{EOM3p1}) does not lead
to any significant improvement but rather introduces instabilities at
large $U$ (data not shown). Another alternative is the EOM scheme for
the effective fermion model derived in section
\ref{Sec_EFM}. The potential advantage of this approach is that its 
starting point is the opposite limit of large $U$ where the local
Hilbert space can be limited to $3$ states only. This approach does,
however, have another problem: holons and doublons are allowed to
occupy the same site at the same time because the hard-core constraint
is an interaction between these particles which cannot be fully
treated. For $U=12$ and $\gamma=0.025$ we find that the error induced
in the current and density profiles by these unphysical states makes
the results of the effective fermion model approach quantitatively
unreliable, see Fig.~\ref{Fig_HF_vs_EFM}.
Nevertheless, these results show some interesting features which are
not present in the Hartree-Fock approach. As the main qualitative
difference we note that in the effective fermion model a doublon peak
with local densities $\langle n_j\rangle >1$ is clearly visible in the
density profile. In the LCRG data such a doublon contribution also
exists, but is much smaller. We have already shown in
Fig.~\ref{Fig_density_short_times} that this contribution can be seen
very clearly numerically at larger $\gamma$.

The local current operator, Eq.~\eqref{current}, in the effective
fermion model is given by
\begin{equation}
\label{current_EFM}
\mathcal{J}_j = 2 \mathcal{J}_j^\uparrow + \mathcal{J}_j^\downarrow 
\end{equation}
with the doublon and holon currents
$\mathcal{J}_j^{\sigma}=-\im\,\sigma J(c_{j,\sigma}^\dagger
c_{j+1,\sigma} - c_{j+1,\sigma}^\dagger c_{j,\sigma})$ with
$\sigma=\,\uparrow,\downarrow\,=+,-$. The Hamiltonian
\eqref{EFM3} of the effective fermionic model does contain
doublon-holon pair creation and annihilation terms. In a short-time
density profile, holes are spread in a light cone around the
dissipative site. If a doublon excitation is now created on top of
this profile it will have an enhanced probability to recombine with a
hole if it travels towards the dissipative site while traveling away
from the defect it has a higher probability to survive and travel on
ballistically. We therefore expect that a {\it positive} doublon
current $\langle\mathcal{J}_j^\uparrow\rangle$ is associated with the
doublon peak seen in the density profiles in regions where the holon
density is low. Such positive local currents are indeed seen in the
numerical solutions of the EOM's for the effective fermion model, see
Fig.~\ref{Fig_HF_vs_EFM}(c,d). The EFM therefore seems to be able to
qualitatively explain the onset of local current reversals far from
the dissipative site: while the local currents close to the defect are
dominated by the holon current $\mathcal{J}_j^\downarrow$, the faster
doublon excitation can move ballistically on a background without
holes outside the holon light cone leading to small local currents
which are positive. This is---in a loose sense---reminiscent of
Hawking radiation where particle-antiparticle pairs are created close
to the event horizon with one particle falling back into the black
hole while the other escapes. In our system there is, however, no
sharp horizon between the region of reduced density and the 'vacuum'
($\langle n_j\rangle=1$) and the energies of the escaping doublons
will not show a thermal distribution. The system is not a sonic analog
of a gravitational black hole. We further note that the EFM model is
not able to describe the metastable state in which all local currents
are reversed. It also always overestimates the doublon contribution
because the doublons can travel on top of the holons in the
approximation considered here.

\subsection{Particle loss}
The particles lost at the defect can be detected, for example, in a
cold gas experiment where an electron beam is used to ionize
atoms. The ions then leave the trap and are collected by a
detector.\cite{BarontiniLabouvie} In the QT approach each quantum jump
corresponds to a particle which is removed from the system. The total
number of particles lost at a given time $t$ can therefore be
calculated by counting the quantum jumps,
\begin{equation}
\label{Noft}
N(t)=\lim_{Q\to\infty}\frac{1}{Q}\sum_{i=1}^Q \int _0 ^t  \delta(t'-t_{\mathrm{jump}_i}) dt',
\end{equation}
where $Q$ is the number of QT's. Experimentally, the quantum jumps for
a single QT correspond to a possible timeline of detection events. The
average particle loss rate $\dot{N}(t)$ has to become a constant in
the NESS. For an infinite system, it is important to distinguish the
particle loss, Eq.~\eqref{Noft}, measured by a detector from the
overall change of the density profile
\begin{equation}
\label{Doft}
D(t) = \sum_{l=-\infty}^{\infty} n_l(t=0) - n_l(t),
\end{equation}
seen, for example, by in-situ imaging. While $D(t)$ and $N(t)$ are
identical for a finite system, this is no longer the case if an
infinite reservoir of particles exists. For the infinite system we
expect a non-trivial time-independent density profile
$D_\infty=D(t\to\infty)$ in the NESS.

\begin{figure}[!ht]
\includegraphics*[width=0.99\columnwidth]{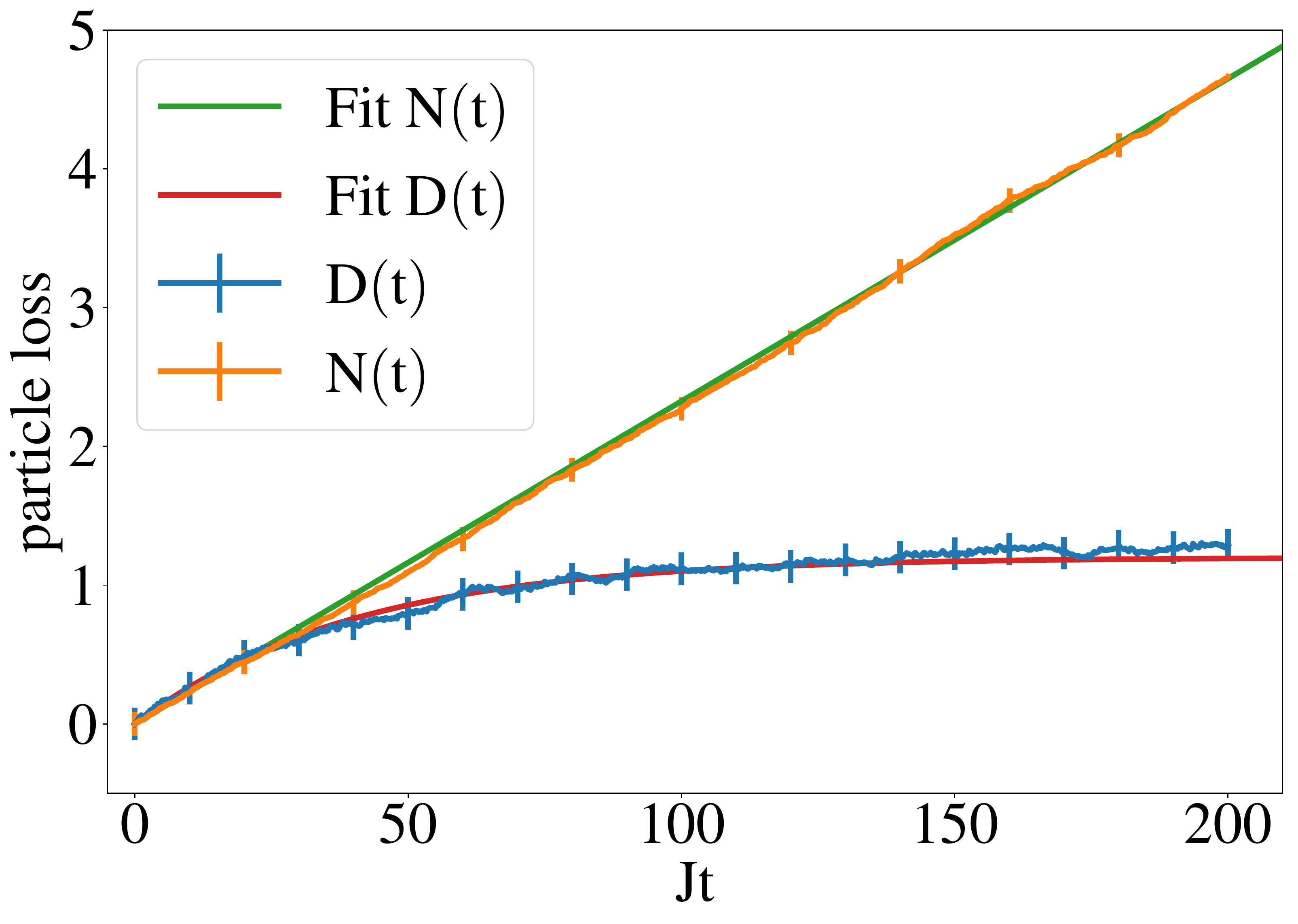}
\caption{Particle loss $N(t)$ and overall change of the density profile 
$D(t)$ for $U=12$ and $\gamma=0.025$. The lines are fits, see text.} 
\label{Fig_part_loss}
\end{figure}

In Fig.~\ref{Fig_part_loss} we show LCRG results for $N(t)$ and
$D(t)$.
At short times $D(t)\approx N(t)$ but at times $t\gtrsim 20$ both
start to deviate. Note that this is roughly the time scale where the
region of reduced particle density stops to extend ballistically with
the holon velocity $v\sim 2J$, see
Fig.~\ref{Fig_density_short_times}. The change in the density profile
can be well approximated by
\begin{equation}
\label{Doft3}
 D(t) \sim A(1-\mathrm{e}^{-\gamma t }),
\end{equation}
with $\gamma$ being the dissipation rate as has also been observed
previously in Ref.~\onlinecite{Barmettler2011}. This seems to suggest
that the time scale for reaching the steady state is $\sim
J/\gamma$. It is important to stress once more, that this is not the
case. Our simulations show that the density profile continues to
change substantially for times $t\gg J/\gamma$. The continuing density
reduction at sites close to the defect is, however, largely
compensated for by a refilling of sites further away from the defect,
making $D(t)$ almost constant for $t>J/\gamma$. The short time
expansion, $D(t)\approx A\gamma t$ does not only capture the behavior
of $D(t)$ at times $t\lesssim 20$ but also yields a good approximation
for the particle loss rate $\dot N(t)\sim A\gamma$. Within error bars,
$\dot N(t)$ does not change as a function of time and is therefore not
a useful quantity to detect whether or not the NESS has been reached.

\begin{figure}[!ht]
\includegraphics*[width=0.99\linewidth]{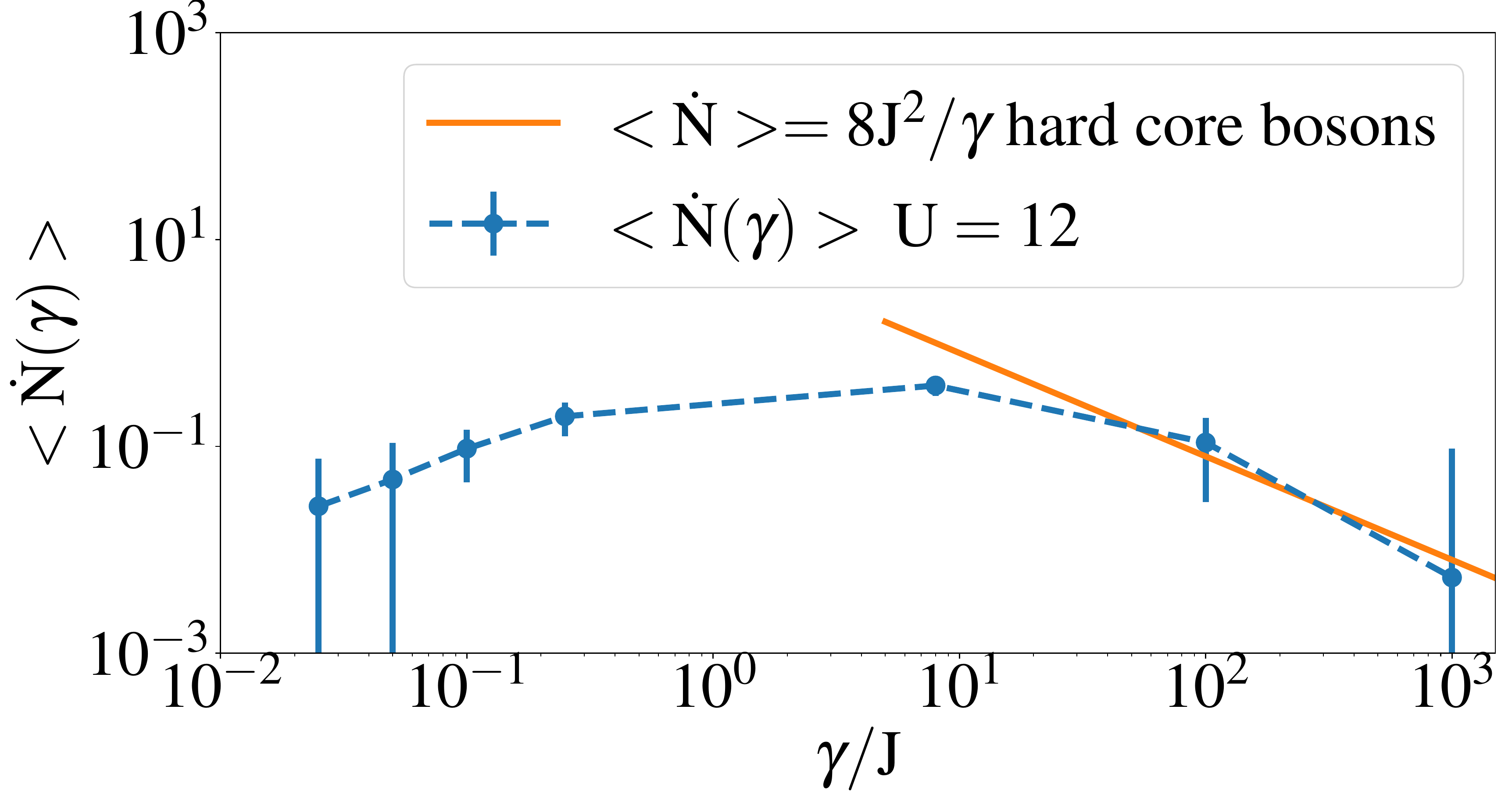}
\caption{Particle loss rate $\dot{N}(\gamma,t\gg J/\gamma)$ for $U=12$ 
as a function of dissipation strength $\gamma$. The error bars denote
the corresponding statistical errors. For large dissipation, the
numerical results are well described by the perturbative result $\dot
N(\gamma,t\gg J/\gamma)=8\gamma^2/J$.}
\label{Fig_loss_rate}
\end{figure}

In Fig.~\ref{Fig_loss_rate} the constant rate $\dot{N}(\gamma,t\gg
J/\gamma)$ is shown as a function of the dissipation strength
$\gamma$.
The loss rate goes through a maximum at $\gamma/J\approx 8$ before
falling of $\sim\gamma^2/J$ for large dissipation
strengths.\cite{Garcia-Ripoll2009} This counterintuitive effect is
known as quantum Zeno dynamics. Large dissipation strengths
effectively stabilize configurations at long times where the
dissipative site is unoccupied with $\gamma\gg J$ effectively acting
as a potential barrier strongly reducing the hopping onto the lossy
site. Our results for the loss rate are consistent with previous
numerical and experimental
studies.\cite{Barmettler2011,BarontiniLabouvie,Garcia-Ripoll2009}

\subsection{Long-range correlations}
The quench dynamics we are investigating here starts from a ground
state deep inside the Mott insulating phase. This state has
exponentially decaying density-density correlations with a rather
small correlation length of about half a lattice site, see
Fig.~\ref{Fig_GS_corr}. Here we want to study how these correlations
change once the dissipative dynamics sets in. We concentrate on the
connected equal time density-density correlation function between the
dissipative site and other sites in the lattice, see Eq.~\eqref{gc}.

\begin{figure}[!ht]
\includegraphics*[width=0.99\columnwidth]{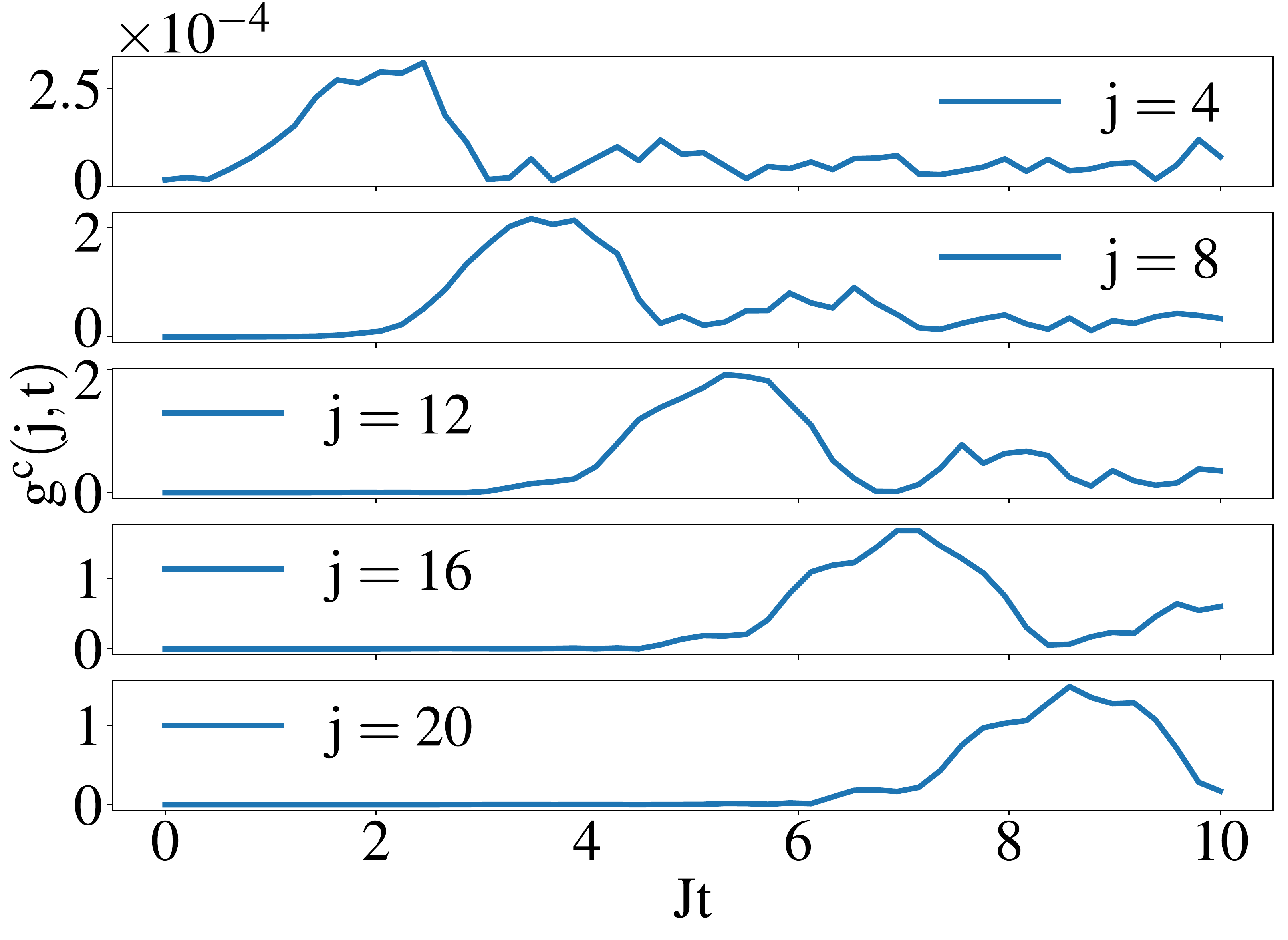}
\caption{QT-LCRG data for the time-evolution of $g^c(j,t)$ at selected sites of the chain 
for $U=12$ and $\gamma=0.025$. A density wave propagates through the
chain with velocity $v \approx 2J$.} 
\label{Fig_dens_pert}
\end{figure}

\begin{figure}[!ht]
\includegraphics*[width=0.99\linewidth]{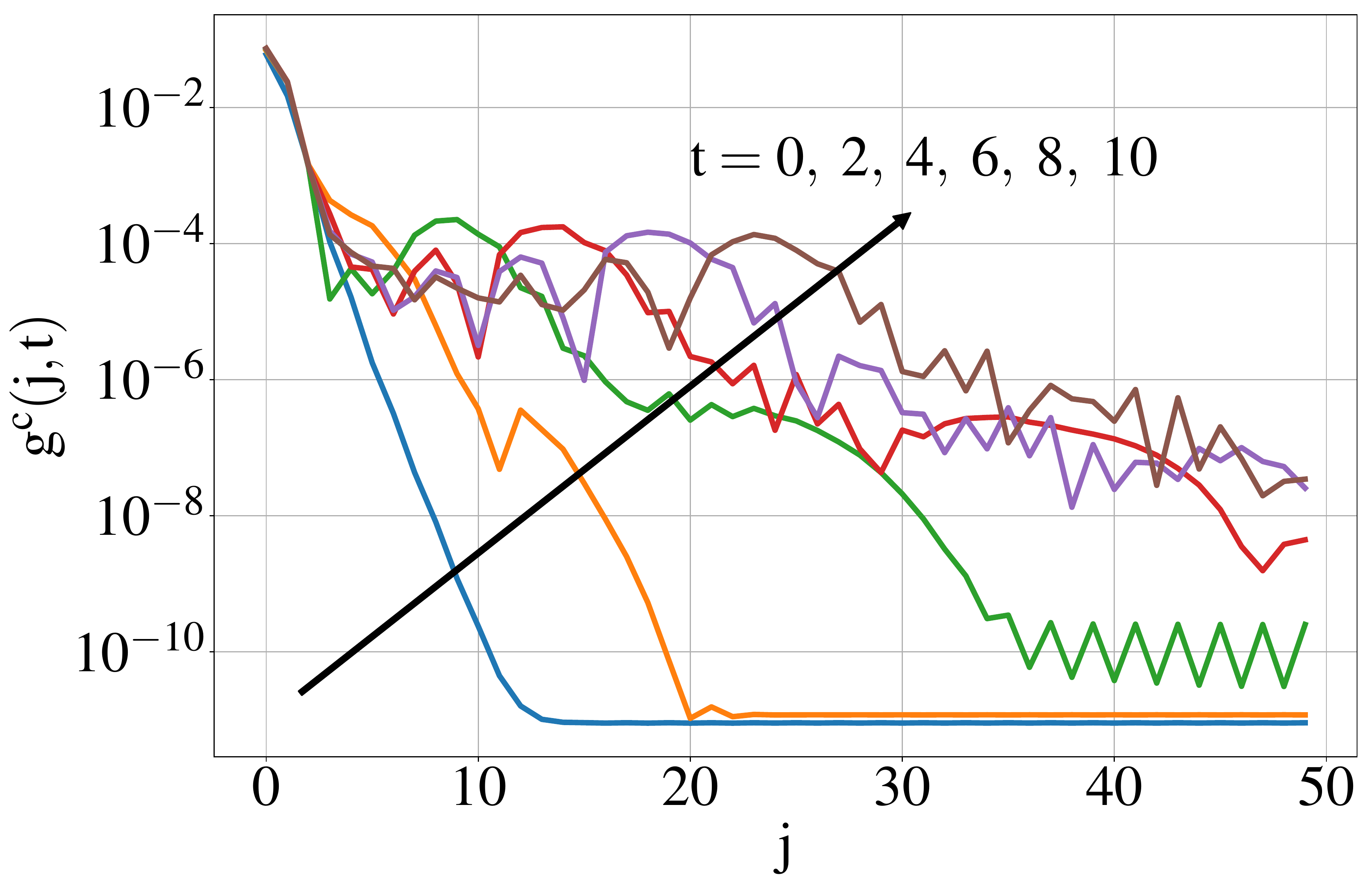}
\caption{The spatial profile of $g^c(j,t)$  at selected times starting from a 
Mott-insulating ground state at $U=12$. At time $t=0$ the correlations
decay exponentially with $\xi\approx 0.57$. The weak dissipative
defect induces correlations inside the holon light cone which are much
longer ranged. The maximal error of the data is of the order $\sim
10^{-6}$.}
\label{Fig_dens_corr}
\end{figure}

At short times, the density perturbation created by turning on the
dissipative process at site $j=0$ at time $t=0$ travels with the holon
velocity $v\sim 2J$ through the system creating a light cone, see
Fig.~\ref{Fig_dens_pert}.
While this density wave travels through the chain, it leaves behind
sites which are stronger correlated than in the initial
state. Fig.~\ref{Fig_dens_corr} shows that the correlations between
the lossy site and sites inside the holon light cone for times $\leq
10$ even appear to be long-ranged.
Obtaining accurate data for the density-density correlation function
requires to calculate $\sim 10000$ QT's which is computationally very
demanding. The data in Fig.~\ref{Fig_dens_corr} are therefore limited
to short times. Based on these data it is impossible to analyze in
more detail if truly long-ranged, power-law decaying, or exponentially
decaying correlations with a large correlation length are
established. 

\subsection{Initial conditions and NESS}
In Ref.~\onlinecite{Labouvie2016} a BHM with local particle loss was
studied. The experiment showed a bistability in a certain parameter
regime: different steady states are reached depending on whether or
not the lossy site has the same filling as the other sites or is empty
in the initial state. In contrast to our study, the experiment was
performed in the superfluid regime with each site occupied on average
by several hundred bosons.

\begin{figure}[!ht]
\includegraphics*[width=0.99\linewidth]{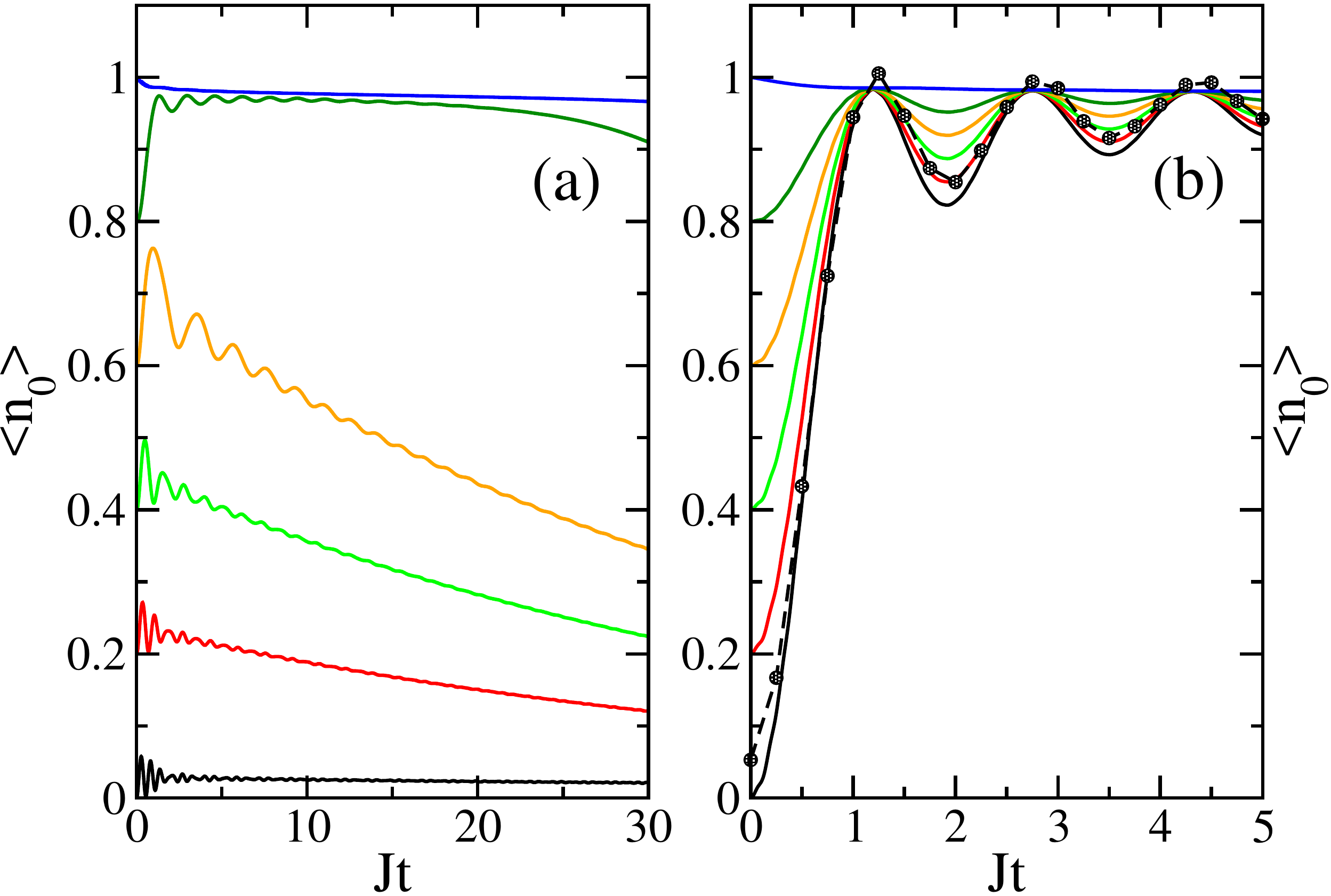}
\caption{Time evolution of the density at the lossy site for initial states with different 
fillings of the lossy site for $U=12$ and $\gamma=0.025$. (a) Results
of the Hartree-Fock EOM approximation do show a bistability. (b) No
bistability is seen using the EOM's for the EFM. The result for
initial filling $\langle n_0\rangle\approx 0$ is in good agreement
with the QT-LCRG data (dots).}
\label{Fig_EOM_instab}
\end{figure}

In the following we will investigate if a similar bistability also
exists deep in the Mott insulating phase. Similar to the experiment,
we modify the density $\langle n_0\rangle$ in the Mott-insulating
initial state. In Fig.~\ref{Fig_EOM_instab}(a) we show results for the
evolution of $\langle n_0(t)\rangle$ obtained using the Hartree-Fock
EOM aproximation for initial states with densities $\langle
n(0)\rangle \in [0,1]$.
Interestingly, the results indeed point to a bistability where the
site $n_0$ becomes almost completely empty or refills almost
completely at intermediate times with a critical filling $\langle
n_0^{\textrm{crit}}(0)\rangle\sim 0.7$. The results obtained using the
EOM's for the effective fermion model are shown in
Fig.~\ref{Fig_EOM_instab}(b) and are very different from the
Hartree-Fock approximation. For all initial states the site $n_0$
fills up again over a rather short time scale. There is no
bistability. The results for initial filling $\langle n_0\rangle=0$
for the effective fermion model are consistent with QT-LCRG data, see
symbols in Fig.~\ref{Fig_EOM_instab}(b). Note that the initial states
in the two approaches are slightly different: We solve the EOM's for
an initial product state with $\langle n_0\rangle =0$ and $\langle
n_j\rangle =1$ for all other sites. In the QT-LCRG calculations, on
the other hand, we first calculate the ground state $|\Psi_0\rangle$
at $U=12$. We then obtain the initial state as
$b_0|\Psi_0\rangle$. Because $|\Psi_0\rangle$ is not a product state,
part of the density is removed from neighboring sites and $\langle
n_0\rangle \approx 0.05$ in the initial state.

While the Hartree-Fock and the EFM approach yield similar results at
short times if we start from the initial state with $\langle
n_j\rangle =1$ at all sites, only the EFM approach is able to describe
the short-time dynamics properly if we start with a reduced density at
the lossy site. This underlines that the EFM approach does capture the
essential aspects of the short-time dynamics and is a good basis to
qualitatively understand the properties of the system deep in the
Mott-insulating phase. We also note that we have only studied one
particular loss rate, $\gamma=0.025$, here. Investigating whether or
not bistabilities do occur in the Mott-insulating initial state for
larger loss rates is beyond the scope of this study.

\subsection{Entanglement entropy}
The QT-LCRG algorithm is based on approximating the time-evolved state
as a matrix product. The success of such an approach hinges on the
amount of entanglement entropy produced by the time evolution. The
Hilbert space is truncated using a reduced density matrix
$\rho_{\textrm{red}} = \tr_E \rho$ where $\rho$ is the full density
matrix and $E$ the part of the system which is traced out. The
entanglement entropy is then given by
\begin{equation}
\label{Sent}
S_{\textrm{ent}} = -\tr \{\rho_{\textrm{red}}\ln\rho_{\textrm{red}}\}
\end{equation}
and is bounded by $\ln\chi$ where $\chi$ is the dimension of
$\rho_{\textrm{red}}$. Since the matrix dimensions which can be
handled numerically is limited in practice, only states with
$S_{\textrm{ent}}\ll\ln\chi$ can be faithfully
represented.\cite{Verstraete2006} It is therefore interesting to study
the time evolution of the entanglement entropy for the lossy BHM.

\begin{figure}[!ht]
\includegraphics*[width=0.99\linewidth]{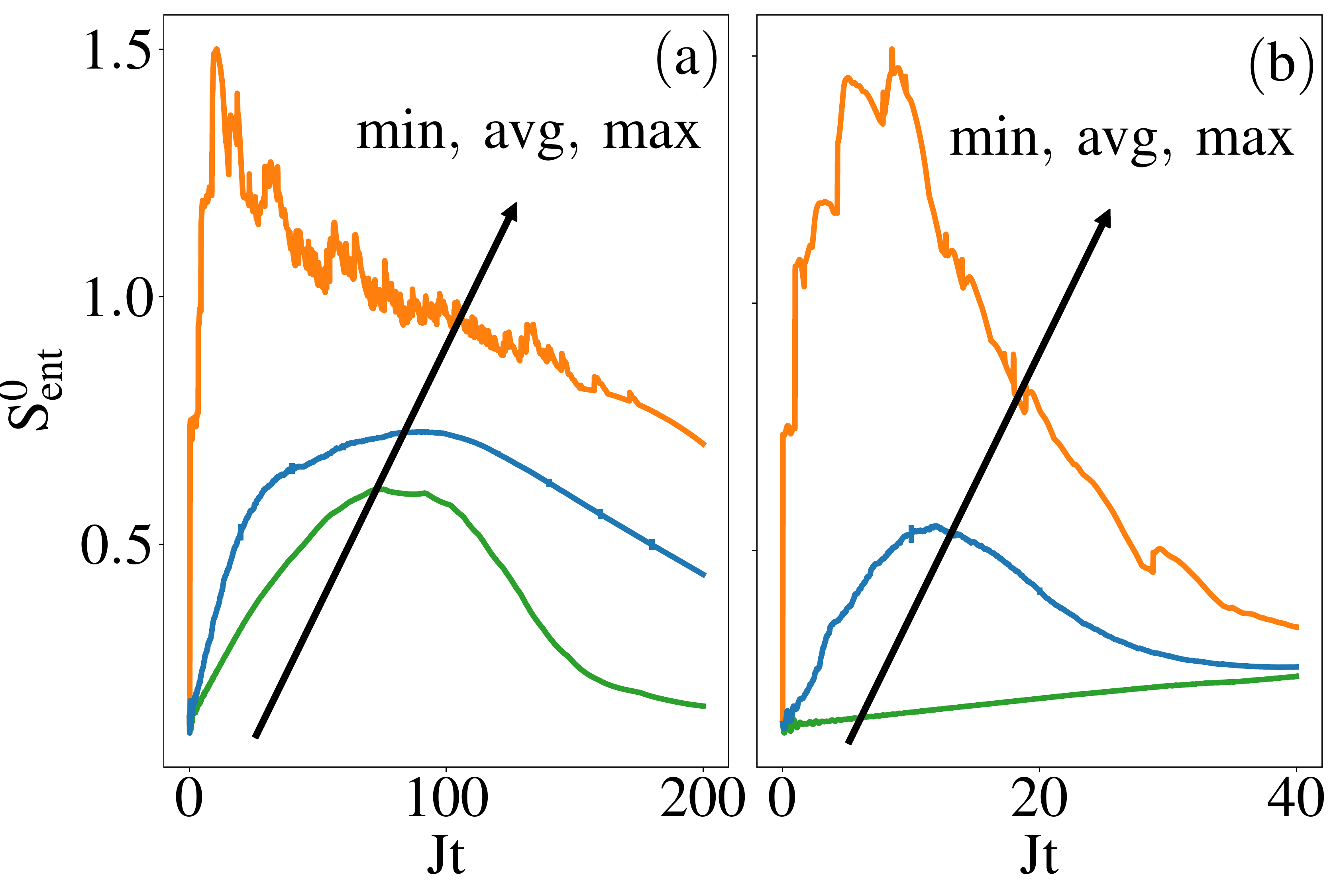}
\caption{Entanglement entropy $S^0_{\textrm{ent}}$ for $U=12$ and (a) $\gamma=0.025$, (b) $\gamma=0.1$.}
\label{Fig_entropies}
\end{figure}

In the QT approach, $S_{\textrm{ent}}(t)$ will be different for each
trajectory. In order to simulate the time evolution we have to keep a
sufficient number of states $\chi$ such that the entropy for the QT's
with the most entanglement always remains small compared to
$\ln\chi$. In the following, we will concentrate on the entanglement
entropy $S^0_{\textrm{ent}}$ obtained by tracing out all sites in the
density matrix to the right of the lossy site $j=0$. Note that the
system is not translationally invariant. In the algorithm we also need
the reduced density matrix where all sites to the right of $j=1$ are
traced out. For small loss rates the entropies for both matrices are,
however, similar so that it suffices to consider $S^0_{\textrm{ent}}$
here. In Fig.~\ref{Fig_entropies} the maximal, minimal, and the
entropy averaged over all QT's are shown for $\gamma=0.025$ and
$\gamma=0.1$.
Quantum jumps typically lead to an abrupt increase of the entanglement
entropy. Trajectories therefore exist which already have significant
entanglement at short times. The imaginary part of the effective
Hamiltonian, on the other hand, suppresses entanglement. At short and
intermediate times, the trajectory with the minimal entanglement is
the one which contains no quantum jumps, while the trajectory with the
maximal entanglement contains many jumps. At long times the two curves
for the extrema apparently converge, pointing to a NESS or metastable
state which has low entanglement. The average entanglement entropy
first shows an approximately logarithmic growth, reaches a maximum,
and then starts dropping almost linearly. For the simulations this
means that one has to keep sufficient states to overcome the maximum
in the entanglement entropy at intermediate times. The simulation time
is then not restricted by a growing $S^0_{\textrm{ent}}$---as is
typically the case for quenches in closed quantum systems---but rather
by the number of renormalization group steps which can be performed
before the accumulated truncation error leads to a breakdown. For
small dissipation rates the QT-LCRG is therefore an attractive tool to
investigate the long-time dynamics of infinite one-dimensional quantum
systems.

\section{Conclusions}
\label{Concl}
Using a novel algorithm which combines the quantum trajectory approach
with the lightcone renormalization group, we have investigated the
dynamics of the Bose-Hubbard model at long-times, $t\gg
J/\gamma$. Starting from a Mott-insulating initial state we found that
for weak particle loss, $\gamma \ll J \ll U$, an intriguing long-time
dynamics takes place. 

Counterintuitively, a reversal of local currents sets in at times $t
\sim J/\gamma$ leading to a state where almost all local currents are
equal and pointing {\it away} from the lossy site. We argued that this
state cannot be the steady but rather is an unusual metastable
state. In the steady state, all currents are equal and are pointing
{\it towards} the lossy site. The currents therefore have to reverse
again at longer times. The reversal of local currents at intermediate
times at sites outside the region with substantially reduced particle
density can be qualitatively understood in an effective fermion
description. In this approach the local Hilbert space is restricted to
three states: empty (holon), singly occupied (vacuum), and doubly
occupied (doublon). While fermionizing the model discards unphysical
states with more than one holon or more than one doublon per site,
doublon and holon can still occupy the same site. While these
unphysical states mean that this approach is quantitatively not fully
reliable, it does explain the qualitative features of the dynamics
seen in the numerical simulations. The Bose-Hubbard Hamiltonian in the
effective fermion representation contains terms annihilating or
creating holon-doublon pairs on neighboring sites. For a pair created
near the edge of the region with reduced density, in particular, the
holon has an adhanced probability to move towards the lossy site while
the doublon is more likely to escape. The process is---in a loose
sense---reminiscent of Hawking radiation near the event horizon of a
black hole and leads to local currents which are pointing away from
the dissipative site.

For a system with infinite particle reservoirs as considered here it
is important to distinguish between the density loss as measured by a
detector and the overall change of the density profile. While the
former is a linear function of time with constant slope for all times
and is therefore not useful to detect whether or not the system has
reached the steady state, the latter exponentially slowly approaches a
constant with a rate $\sim\gamma$. However, this does not imply that
the steady state is reached on time scales $\sim J/\gamma$. The
density profile continues to change substantially at times
$t>J/\gamma$ with a density loss at sites close to the defect almost
compensated for by a refilling of sites further away from the defect.

In the density-density correlations at short times a light cone
structure is clearly visible. Once the front of the light cone has
passed, correlations between the dissipative site and sites inside the
cone are established which are much longer ranged than in the initial
Mott-insulating state. An interesting question is if the dissipative
dynamics can create truly long-range correlations. Numerically, this
question is very difficult to address because a large number of
trajectories are required to obtain reliable results for two-point
correlation functions. Based on the data for times $Jt<10$ we cannot
decide if the correlations are truly long-ranged, power-law decaying,
or even exponentially decaying with a very long correlation length.

Starting from initial states with different initial filling of the
dissipative site we studied if the Lindblad dynamics can lead to
different steady or metastable states. While the Hartree-Fock equation
of motion approach suggests a bistability, similar to the one seen in
a cold gas experiment in the superfluid regime, such a behavior is not
confirmed in the effective fermion model. For the small dissipation
rate $\gamma$ considered, all initial states with different filling of
the lossy site seem to lead to the same steady state. We showed, in
particular, that the numerically calculated dynamics starting from the
state where the lossy site is initially empty is in good agreement
with the effective fermion model result. This underlines that the
effective fermion model is a useful approach to understand the
qualitative features of the open Bose-Hubbard dynamics at short and
intermediate times deep inside the Mott-insulating regime.

The chosen model and parameters can be realized in a cold gas
experiment. Detecting the doublons moving away from the dissipative
site would be an indicator for the separation of holon-doublon pairs
by the dissipation. While the considered system does not have a sharp
event horizon, studying particles expelled from the dissipative region
might be a step towards realizing sonic analogs of gravitational black
holes. In order to achieve a full analogy, local losses in Bose gases
in the superfluid phase in higher dimensions need to be
realized.\cite{GarayAnglin} Such systems are, however, more difficult
to analyze theoretically beyond the mean-field level so that a careful
study of losses in one-dimensional lattice models might be a useful
interim goal.

\acknowledgments
We acknowledge support by the Natural Sciences and Engineering
Research Council (NSERC, Canada) and by the Deutsche
Forschungsgemeinschaft (DFG) via Research Unit FOR 2316. We are
grateful for the computing resources provided by Compute Canada and
Westgrid as well as for the GPU unit made available by
NVIDIA. J.S.~acknowledges helpful discussions with J.~Anglin,
M.~Fleischhauer, and H.~Ott.


\begin{thebibliography}{42}
\expandafter\ifx\csname natexlab\endcsname\relax\def\natexlab#1{#1}\fi
\expandafter\ifx\csname bibnamefont\endcsname\relax
  \def\bibnamefont#1{#1}\fi
\expandafter\ifx\csname bibfnamefont\endcsname\relax
  \def\bibfnamefont#1{#1}\fi
\expandafter\ifx\csname citenamefont\endcsname\relax
  \def\citenamefont#1{#1}\fi
\expandafter\ifx\csname url\endcsname\relax
  \def\url#1{\texttt{#1}}\fi
\expandafter\ifx\csname urlprefix\endcsname\relax\def\urlprefix{URL }\fi
\providecommand{\bibinfo}[2]{#2}
\providecommand{\eprint}[2][]{\url{#2}}

\bibitem[{\citenamefont{Bloch}(2005)}]{Bloch_NatPhys}
\bibinfo{author}{\bibfnamefont{I.}~\bibnamefont{Bloch}}, \bibinfo{journal}{Nat.
  Phys.} \textbf{\bibinfo{volume}{1}}, \bibinfo{pages}{23}
  (\bibinfo{year}{2005}).

\bibitem[{\citenamefont{{Tomita} et~al.}(2017)\citenamefont{{Tomita},
  {Nakajima}, {Danshita}, {Takasu}, and {Takahashi}}}]{TomitaNakajima}
\bibinfo{author}{\bibfnamefont{T.}~\bibnamefont{{Tomita}}},
  \bibinfo{author}{\bibfnamefont{S.}~\bibnamefont{{Nakajima}}},
  \bibinfo{author}{\bibfnamefont{I.}~\bibnamefont{{Danshita}}},
  \bibinfo{author}{\bibfnamefont{Y.}~\bibnamefont{{Takasu}}}, \bibnamefont{and}
  \bibinfo{author}{\bibfnamefont{Y.}~\bibnamefont{{Takahashi}}},
  \bibinfo{journal}{arXiv:1705.09942}  (\bibinfo{year}{2017}).

\bibitem[{\citenamefont{Daley}(2014)}]{Daley2014}
\bibinfo{author}{\bibfnamefont{A.~J.} \bibnamefont{Daley}},
  \bibinfo{journal}{Advances in Physics} \textbf{\bibinfo{volume}{63}},
  \bibinfo{pages}{77} (\bibinfo{year}{2014}).

\bibitem[{\citenamefont{Verstraete et~al.}(2009)\citenamefont{Verstraete, Wolf,
  and Ignacio~Cirac}}]{Verstraete2009b}
\bibinfo{author}{\bibfnamefont{F.}~\bibnamefont{Verstraete}},
  \bibinfo{author}{\bibfnamefont{M.~M.} \bibnamefont{Wolf}}, \bibnamefont{and}
  \bibinfo{author}{\bibfnamefont{J.}~\bibnamefont{Ignacio~Cirac}},
  \bibinfo{journal}{Nat. Phys.} \textbf{\bibinfo{volume}{5}},
  \bibinfo{pages}{633} (\bibinfo{year}{2009}).

\bibitem[{\citenamefont{Syassen et~al.}(2008)\citenamefont{Syassen, Bauer,
  Lettner, Volz, Dietze, Garc{\'\i}a-Ripoll, Cirac, Rempe, and
  D{\"u}rr}}]{Syassen2008}
\bibinfo{author}{\bibfnamefont{N.}~\bibnamefont{Syassen}},
  \bibinfo{author}{\bibfnamefont{D.~M.} \bibnamefont{Bauer}},
  \bibinfo{author}{\bibfnamefont{M.}~\bibnamefont{Lettner}},
  \bibinfo{author}{\bibfnamefont{T.}~\bibnamefont{Volz}},
  \bibinfo{author}{\bibfnamefont{D.}~\bibnamefont{Dietze}},
  \bibinfo{author}{\bibfnamefont{J.~J.} \bibnamefont{Garc{\'\i}a-Ripoll}},
  \bibinfo{author}{\bibfnamefont{J.~I.} \bibnamefont{Cirac}},
  \bibinfo{author}{\bibfnamefont{G.}~\bibnamefont{Rempe}}, \bibnamefont{and}
  \bibinfo{author}{\bibfnamefont{S.}~\bibnamefont{D{\"u}rr}},
  \bibinfo{journal}{Science} \textbf{\bibinfo{volume}{320}},
  \bibinfo{pages}{1329} (\bibinfo{year}{2008}).

\bibitem[{\citenamefont{Barontini et~al.}(2013)\citenamefont{Barontini,
  Labouvie, Stubenrauch, Vogler, Guarrera, and Ott}}]{BarontiniLabouvie}
\bibinfo{author}{\bibfnamefont{G.}~\bibnamefont{Barontini}},
  \bibinfo{author}{\bibfnamefont{R.}~\bibnamefont{Labouvie}},
  \bibinfo{author}{\bibfnamefont{F.}~\bibnamefont{Stubenrauch}},
  \bibinfo{author}{\bibfnamefont{A.}~\bibnamefont{Vogler}},
  \bibinfo{author}{\bibfnamefont{V.}~\bibnamefont{Guarrera}}, \bibnamefont{and}
  \bibinfo{author}{\bibfnamefont{H.}~\bibnamefont{Ott}},
  \bibinfo{journal}{Phys. Rev. Lett.} \textbf{\bibinfo{volume}{110}},
  \bibinfo{pages}{035302} (\bibinfo{year}{2013}).

\bibitem[{\citenamefont{Labouvie et~al.}(2015)\citenamefont{Labouvie, Santra,
  Heun, Wimberger, and Ott}}]{Labouvie2015a}
\bibinfo{author}{\bibfnamefont{R.}~\bibnamefont{Labouvie}},
  \bibinfo{author}{\bibfnamefont{B.}~\bibnamefont{Santra}},
  \bibinfo{author}{\bibfnamefont{S.}~\bibnamefont{Heun}},
  \bibinfo{author}{\bibfnamefont{S.}~\bibnamefont{Wimberger}},
  \bibnamefont{and} \bibinfo{author}{\bibfnamefont{H.}~\bibnamefont{Ott}},
  \bibinfo{journal}{Phys. Rev. Lett.} \textbf{\bibinfo{volume}{115}},
  \bibinfo{pages}{050601} (\bibinfo{year}{2015}).

\bibitem[{\citenamefont{Labouvie et~al.}(2016)\citenamefont{Labouvie, Santra,
  Heun, and Ott}}]{Labouvie2016}
\bibinfo{author}{\bibfnamefont{R.}~\bibnamefont{Labouvie}},
  \bibinfo{author}{\bibfnamefont{B.}~\bibnamefont{Santra}},
  \bibinfo{author}{\bibfnamefont{S.}~\bibnamefont{Heun}}, \bibnamefont{and}
  \bibinfo{author}{\bibfnamefont{H.}~\bibnamefont{Ott}},
  \bibinfo{journal}{Phys. Rev. Lett.} \textbf{\bibinfo{volume}{116}},
  \bibinfo{pages}{235302} (\bibinfo{year}{2016}).

\bibitem[{\citenamefont{Kepesidis and Hartmann}(2012)}]{Kepesidis2012}
\bibinfo{author}{\bibfnamefont{K.~V.} \bibnamefont{Kepesidis}}
  \bibnamefont{and} \bibinfo{author}{\bibfnamefont{M.~J.}
  \bibnamefont{Hartmann}}, \bibinfo{journal}{Phys. Rev. A}
  \textbf{\bibinfo{volume}{85}}, \bibinfo{pages}{063620}
  (\bibinfo{year}{2012}).

\bibitem[{\citenamefont{Barmettler and Kollath}(2011)}]{Barmettler2011}
\bibinfo{author}{\bibfnamefont{P.}~\bibnamefont{Barmettler}} \bibnamefont{and}
  \bibinfo{author}{\bibfnamefont{C.}~\bibnamefont{Kollath}},
  \bibinfo{journal}{Phys. Rev. A} \textbf{\bibinfo{volume}{84}},
  \bibinfo{pages}{041606} (\bibinfo{year}{2011}).

\bibitem[{\citenamefont{Garc{\'\i}a-Ripoll
  et~al.}(2009)\citenamefont{Garc{\'\i}a-Ripoll, D{\"u}rr, Syassen, Bauer,
  Lettner, Rempe, and Cirac}}]{Garcia-Ripoll2009}
\bibinfo{author}{\bibfnamefont{J.~J.} \bibnamefont{Garc{\'\i}a-Ripoll}},
  \bibinfo{author}{\bibfnamefont{S.}~\bibnamefont{D{\"u}rr}},
  \bibinfo{author}{\bibfnamefont{N.}~\bibnamefont{Syassen}},
  \bibinfo{author}{\bibfnamefont{D.~M.} \bibnamefont{Bauer}},
  \bibinfo{author}{\bibfnamefont{M.}~\bibnamefont{Lettner}},
  \bibinfo{author}{\bibfnamefont{G.}~\bibnamefont{Rempe}}, \bibnamefont{and}
  \bibinfo{author}{\bibfnamefont{J.~I.} \bibnamefont{Cirac}},
  \bibinfo{journal}{New Journal of Physics} \textbf{\bibinfo{volume}{11}},
  \bibinfo{pages}{013053} (\bibinfo{year}{2009}).

\bibitem[{\citenamefont{Daley et~al.}(2009)\citenamefont{Daley, Taylor, Diehl,
  Baranov, and Zoller}}]{Daley2009}
\bibinfo{author}{\bibfnamefont{A.~J.} \bibnamefont{Daley}},
  \bibinfo{author}{\bibfnamefont{J.~M.} \bibnamefont{Taylor}},
  \bibinfo{author}{\bibfnamefont{S.}~\bibnamefont{Diehl}},
  \bibinfo{author}{\bibfnamefont{M.}~\bibnamefont{Baranov}}, \bibnamefont{and}
  \bibinfo{author}{\bibfnamefont{P.}~\bibnamefont{Zoller}},
  \bibinfo{journal}{Phys. Rev. Lett.} \textbf{\bibinfo{volume}{102}},
  \bibinfo{pages}{040402} (\bibinfo{year}{2009}).

\bibitem[{\citenamefont{van Nieuwenburg et~al.}(2017)\citenamefont{van
  Nieuwenburg, Mal, Daley, and Fischer}}]{NieuwenburgMalo}
\bibinfo{author}{\bibfnamefont{E.~P.} \bibnamefont{van Nieuwenburg}},
  \bibinfo{author}{\bibfnamefont{J.~Y.} \bibnamefont{Mal}},
  \bibinfo{author}{\bibfnamefont{A.~J.} \bibnamefont{Daley}}, \bibnamefont{and}
  \bibinfo{author}{\bibfnamefont{M.~H.} \bibnamefont{Fischer}},
  \bibinfo{journal}{arXiv:1706.00788}  (\bibinfo{year}{2017}).

\bibitem[{\citenamefont{Bernier et~al.}(2017)\citenamefont{Bernier, Tan,
  Bonnes, Guo, Poletti, and Kollath}}]{BernierTan}
\bibinfo{author}{\bibfnamefont{J.-S.} \bibnamefont{Bernier}},
  \bibinfo{author}{\bibfnamefont{R.}~\bibnamefont{Tan}},
  \bibinfo{author}{\bibfnamefont{L.}~\bibnamefont{Bonnes}},
  \bibinfo{author}{\bibfnamefont{C.}~\bibnamefont{Guo}},
  \bibinfo{author}{\bibfnamefont{D.}~\bibnamefont{Poletti}}, \bibnamefont{and}
  \bibinfo{author}{\bibfnamefont{C.}~\bibnamefont{Kollath}},
  \bibinfo{journal}{arXiv:1702.04136}  (\bibinfo{year}{2017}).

\bibitem[{\citenamefont{White}(1992)}]{White1992}
\bibinfo{author}{\bibfnamefont{S.~R.} \bibnamefont{White}},
  \bibinfo{journal}{Phys. Rev. Lett.} \textbf{\bibinfo{volume}{69}},
  \bibinfo{pages}{2863} (\bibinfo{year}{1992}).

\bibitem[{\citenamefont{Daley et~al.}(2004)\citenamefont{Daley, Kollath,
  Schollw\"ock, and Vidal}}]{DaleyKollath}
\bibinfo{author}{\bibfnamefont{A.~J.} \bibnamefont{Daley}},
  \bibinfo{author}{\bibfnamefont{C.}~\bibnamefont{Kollath}},
  \bibinfo{author}{\bibfnamefont{U.}~\bibnamefont{Schollw\"ock}},
  \bibnamefont{and} \bibinfo{author}{\bibfnamefont{G.}~\bibnamefont{Vidal}},
  \bibinfo{journal}{J. Stat. Mech.}  \bibinfo{pages}{P04005}
  (\bibinfo{year}{2004}).

\bibitem[{\citenamefont{White and Feiguin}(2004)}]{WhiteFeiguin}
\bibinfo{author}{\bibfnamefont{S.}~\bibnamefont{White}} \bibnamefont{and}
  \bibinfo{author}{\bibfnamefont{A.~E.} \bibnamefont{Feiguin}},
  \bibinfo{journal}{Phys. Rev. Lett.} \textbf{\bibinfo{volume}{93}},
  \bibinfo{pages}{076401} (\bibinfo{year}{2004}).

\bibitem[{\citenamefont{Vidal}(2003)}]{Vidal2003}
\bibinfo{author}{\bibfnamefont{G.}~\bibnamefont{Vidal}},
  \bibinfo{journal}{Phys. Rev. Lett.} \textbf{\bibinfo{volume}{91}},
  \bibinfo{pages}{147902} (\bibinfo{year}{2003}).

\bibitem[{\citenamefont{Zwolak and Vidal}(2004)}]{Zwolak2004}
\bibinfo{author}{\bibfnamefont{M.}~\bibnamefont{Zwolak}} \bibnamefont{and}
  \bibinfo{author}{\bibfnamefont{G.}~\bibnamefont{Vidal}},
  \bibinfo{journal}{Phys. Rev. Lett.} \textbf{\bibinfo{volume}{93}},
  \bibinfo{pages}{207205} (\bibinfo{year}{2004}).

\bibitem[{\citenamefont{Verstraete et~al.}(2004)\citenamefont{Verstraete,
  Garc\'{\i}a-Ripoll, and Cirac}}]{PhysRevLett.93.207204}
\bibinfo{author}{\bibfnamefont{F.}~\bibnamefont{Verstraete}},
  \bibinfo{author}{\bibfnamefont{J.~J.} \bibnamefont{Garc\'{\i}a-Ripoll}},
  \bibnamefont{and} \bibinfo{author}{\bibfnamefont{J.~I.} \bibnamefont{Cirac}},
  \bibinfo{journal}{Phys. Rev. Lett.} \textbf{\bibinfo{volume}{93}},
  \bibinfo{pages}{207204} (\bibinfo{year}{2004}).

\bibitem[{\citenamefont{Macieszczak et~al.}(2016)\citenamefont{Macieszczak,
  Gu\ifmmode \mbox{\c{t}}\else \c{t}\fi{}\ifmmode~\u{a}\else \u{a}\fi{},
  Lesanovsky, and Garrahan}}]{Macieszczak2016}
\bibinfo{author}{\bibfnamefont{K.}~\bibnamefont{Macieszczak}},
  \bibinfo{author}{\bibfnamefont{M.}~\bibnamefont{Gu\ifmmode \mbox{\c{t}}\else
  \c{t}\fi{}\ifmmode~\u{a}\else \u{a}\fi{}}},
  \bibinfo{author}{\bibfnamefont{I.}~\bibnamefont{Lesanovsky}},
  \bibnamefont{and} \bibinfo{author}{\bibfnamefont{J.~P.}
  \bibnamefont{Garrahan}}, \bibinfo{journal}{Phys. Rev. Lett.}
  \textbf{\bibinfo{volume}{116}}, \bibinfo{pages}{240404}
  (\bibinfo{year}{2016}).

\bibitem[{\citenamefont{M\o{}lmer}(1993)}]{Molmer1993}
\bibinfo{author}{\bibfnamefont{K.}~\bibnamefont{M\o{}lmer}},
  \textbf{\bibinfo{volume}{10}}, \bibinfo{pages}{524} (\bibinfo{year}{1993}).

\bibitem[{\citenamefont{Carmichael}(1993)}]{Carmichael1993}
\bibinfo{author}{\bibfnamefont{H.~J.} \bibnamefont{Carmichael}},
  \bibinfo{journal}{Springer, Berlin}  (\bibinfo{year}{1993}).

\bibitem[{\citenamefont{Breuer and Petruccione}(2002)}]{Theoopenqusys2002}
\bibinfo{author}{\bibfnamefont{H.-P.} \bibnamefont{Breuer}} \bibnamefont{and}
  \bibinfo{author}{\bibfnamefont{F.}~\bibnamefont{Petruccione}},
  \bibinfo{journal}{Oxford University Press}  (\bibinfo{year}{2002}).

\bibitem[{\citenamefont{Enss and Sirker}(2012)}]{EnssSirker}
\bibinfo{author}{\bibfnamefont{T.}~\bibnamefont{Enss}} \bibnamefont{and}
  \bibinfo{author}{\bibfnamefont{J.}~\bibnamefont{Sirker}},
  \bibinfo{journal}{New J. Phys.} \textbf{\bibinfo{volume}{14}},
  \bibinfo{pages}{023008} (\bibinfo{year}{2012}).

\bibitem[{\citenamefont{W\"urtz et~al.}(2009)\citenamefont{W\"urtz, Langen,
  Gericke, Koglbauer, and Ott}}]{WuertzLangen}
\bibinfo{author}{\bibfnamefont{P.}~\bibnamefont{W\"urtz}},
  \bibinfo{author}{\bibfnamefont{T.}~\bibnamefont{Langen}},
  \bibinfo{author}{\bibfnamefont{T.}~\bibnamefont{Gericke}},
  \bibinfo{author}{\bibfnamefont{A.}~\bibnamefont{Koglbauer}},
  \bibnamefont{and} \bibinfo{author}{\bibfnamefont{H.}~\bibnamefont{Ott}},
  \bibinfo{journal}{Phys. Rev. Lett.} \textbf{\bibinfo{volume}{103}},
  \bibinfo{pages}{080404} (\bibinfo{year}{2009}).

\bibitem[{\citenamefont{Kordas et~al.}(2015{\natexlab{a}})\citenamefont{Kordas,
  Witthaut, Buonsante, Vezzani, Burioni, Karanikas, and
  Wimberger}}]{WimbergerReview}
\bibinfo{author}{\bibfnamefont{G.}~\bibnamefont{Kordas}},
  \bibinfo{author}{\bibfnamefont{D.}~\bibnamefont{Witthaut}},
  \bibinfo{author}{\bibfnamefont{P.}~\bibnamefont{Buonsante}},
  \bibinfo{author}{\bibfnamefont{A.}~\bibnamefont{Vezzani}},
  \bibinfo{author}{\bibfnamefont{R.}~\bibnamefont{Burioni}},
  \bibinfo{author}{\bibfnamefont{A.~I.} \bibnamefont{Karanikas}},
  \bibnamefont{and}
  \bibinfo{author}{\bibfnamefont{S.}~\bibnamefont{Wimberger}},
  \bibinfo{journal}{Eur. Phys. J. Special Topics}
  \textbf{\bibinfo{volume}{224}}, \bibinfo{pages}{2127}
  (\bibinfo{year}{2015}{\natexlab{a}}).

\bibitem[{\citenamefont{Kordas et~al.}(2015{\natexlab{b}})\citenamefont{Kordas,
  Witthaut, and Wimberger}}]{KordasWitthaut}
\bibinfo{author}{\bibfnamefont{G.}~\bibnamefont{Kordas}},
  \bibinfo{author}{\bibfnamefont{D.}~\bibnamefont{Witthaut}}, \bibnamefont{and}
  \bibinfo{author}{\bibfnamefont{S.}~\bibnamefont{Wimberger}},
  \bibinfo{journal}{Ann, Phys. (Berlin)} \textbf{\bibinfo{volume}{527}},
  \bibinfo{pages}{619} (\bibinfo{year}{2015}{\natexlab{b}}).

\bibitem[{\citenamefont{Barmettler et~al.}(2012)\citenamefont{Barmettler,
  Poletti, Cheneau, and Kollath}}]{BarmettlerPoletti}
\bibinfo{author}{\bibfnamefont{P.}~\bibnamefont{Barmettler}},
  \bibinfo{author}{\bibfnamefont{D.}~\bibnamefont{Poletti}},
  \bibinfo{author}{\bibfnamefont{M.}~\bibnamefont{Cheneau}}, \bibnamefont{and}
  \bibinfo{author}{\bibfnamefont{C.}~\bibnamefont{Kollath}},
  \bibinfo{journal}{Phys. Rev. A} \textbf{\bibinfo{volume}{85}},
  \bibinfo{pages}{053625} (\bibinfo{year}{2012}).

\bibitem[{\citenamefont{Marc~Cheneau}(2012)}]{CheneauBarmettler}
\bibinfo{author}{\bibnamefont{Marc~Cheneau},
  \bibfnamefont{Peter~Barmettler}~\bibfnamefont{{\it et al.}}}, \bibinfo{journal}{Nature}
  \textbf{\bibinfo{volume}{481}}, \bibinfo{pages}{484} (\bibinfo{year}{2012}).

\bibitem[{\citenamefont{Andraschko and Sirker}(2015)}]{AndraschkoSirker2}
\bibinfo{author}{\bibfnamefont{F.}~\bibnamefont{Andraschko}} \bibnamefont{and}
  \bibinfo{author}{\bibfnamefont{J.}~\bibnamefont{Sirker}},
  \bibinfo{journal}{Phys. Rev. B} \textbf{\bibinfo{volume}{91}},
  \bibinfo{pages}{235132} (\bibinfo{year}{2015}).

\bibitem[{\citenamefont{Bonnes et~al.}(2014)\citenamefont{Bonnes, Charrier, and
  L\"auchli}}]{BonnesCharrier}
\bibinfo{author}{\bibfnamefont{L.}~\bibnamefont{Bonnes}},
  \bibinfo{author}{\bibfnamefont{D.}~\bibnamefont{Charrier}}, \bibnamefont{and}
  \bibinfo{author}{\bibfnamefont{A.~M.} \bibnamefont{L\"auchli}},
  \bibinfo{journal}{Phys. Rev. A} \textbf{\bibinfo{volume}{90}},
  \bibinfo{pages}{033612} (\bibinfo{year}{2014}).

\bibitem[{\citenamefont{Bonnes and L\"auchli}(2014)}]{BonnesLauchli}
\bibinfo{author}{\bibfnamefont{L.}~\bibnamefont{Bonnes}} \bibnamefont{and}
  \bibinfo{author}{\bibfnamefont{A.}~\bibnamefont{L\"auchli}},
  \bibinfo{journal}{arXiv:1411.4831}  (\bibinfo{year}{2014}).

\bibitem[{\citenamefont{Suzuki}(1976)}]{Suzuki1976}
\bibinfo{author}{\bibfnamefont{M.}~\bibnamefont{Suzuki}},
  \bibinfo{journal}{Commun. Math. Phys.}
  \textbf{\bibinfo{volume}{51}}, \bibinfo{pages}{183} (\bibinfo{year}{1976}).

\bibitem[{\citenamefont{Lieb and Robinson}(1972)}]{LiebRobinson}
\bibinfo{author}{\bibfnamefont{E.~H.} \bibnamefont{Lieb}} \bibnamefont{and}
  \bibinfo{author}{\bibfnamefont{D.~W.} \bibnamefont{Robinson}},
  \bibinfo{journal}{Commun. Math. Phys.} \textbf{\bibinfo{volume}{28}},
  \bibinfo{pages}{251} (\bibinfo{year}{1972}).

\bibitem[{\citenamefont{Bravyi et~al.}(2006)\citenamefont{Bravyi, Hastings, and
  Verstraete}}]{BravyiHastings}
\bibinfo{author}{\bibfnamefont{S.}~\bibnamefont{Bravyi}},
  \bibinfo{author}{\bibfnamefont{M.~B.} \bibnamefont{Hastings}},
  \bibnamefont{and}
  \bibinfo{author}{\bibfnamefont{F.}~\bibnamefont{Verstraete}},
  \bibinfo{journal}{Phys. Rev. Lett.} \textbf{\bibinfo{volume}{97}},
  \bibinfo{pages}{050401} (\bibinfo{year}{2006}).

\bibitem[{\citenamefont{Andraschko et~al.}(2014)\citenamefont{Andraschko, Enss,
  and Sirker}}]{AndraschkoEnssSirker}
\bibinfo{author}{\bibfnamefont{F.}~\bibnamefont{Andraschko}},
  \bibinfo{author}{\bibfnamefont{T.}~\bibnamefont{Enss}}, \bibnamefont{and}
  \bibinfo{author}{\bibfnamefont{J.}~\bibnamefont{Sirker}},
  \bibinfo{journal}{Phys. Rev. Lett.} \textbf{\bibinfo{volume}{113}},
  \bibinfo{pages}{217201} (\bibinfo{year}{2014}).

\bibitem[{\citenamefont{Enss et~al.}(2017)\citenamefont{Enss, Andraschko, and
  Sirker}}]{EnssAndraschkoSirker}
\bibinfo{author}{\bibfnamefont{T.}~\bibnamefont{Enss}},
  \bibinfo{author}{\bibfnamefont{F.}~\bibnamefont{Andraschko}},
  \bibnamefont{and} \bibinfo{author}{\bibfnamefont{J.}~\bibnamefont{Sirker}},
  \bibinfo{journal}{Phys. Rev. B} \textbf{\bibinfo{volume}{95}},
  \bibinfo{pages}{045121} (\bibinfo{year}{2017}).

\bibitem[{\citenamefont{Vidal}(2007)}]{VidaliTEBD}
\bibinfo{author}{\bibfnamefont{G.}~\bibnamefont{Vidal}},
  \bibinfo{journal}{Phys. Rev. Lett.} \textbf{\bibinfo{volume}{98}},
  \bibinfo{pages}{070201} (\bibinfo{year}{2007}).

\bibitem[{\citenamefont{Or\'us and Vidal}(2008)}]{Orus2008}
\bibinfo{author}{\bibfnamefont{R.}~\bibnamefont{Or\'us}} \bibnamefont{and}
  \bibinfo{author}{\bibfnamefont{G.}~\bibnamefont{Vidal}},
  \bibinfo{journal}{Phys. Rev. B} \textbf{\bibinfo{volume}{78}},
  \bibinfo{pages}{155117} (\bibinfo{year}{2008}).

\bibitem[{\citenamefont{Verstraete and Cirac}(2006)}]{Verstraete2006}
\bibinfo{author}{\bibfnamefont{F.}~\bibnamefont{Verstraete}} \bibnamefont{and}
  \bibinfo{author}{\bibfnamefont{J.~I.} \bibnamefont{Cirac}},
  \bibinfo{journal}{Phys. Rev. B} \textbf{\bibinfo{volume}{73}},
  \bibinfo{pages}{094423} (\bibinfo{year}{2006}).

\bibitem[{\citenamefont{Garay et~al.}({2000})\citenamefont{Garay, Anglin,
  Cirac, and Zoller}}]{GarayAnglin}
\bibinfo{author}{\bibfnamefont{L.}~\bibnamefont{Garay}},
  \bibinfo{author}{\bibfnamefont{J.}~\bibnamefont{Anglin}},
  \bibinfo{author}{\bibfnamefont{J.}~\bibnamefont{Cirac}}, \bibnamefont{and}
  \bibinfo{author}{\bibfnamefont{P.}~\bibnamefont{Zoller}},
  \bibinfo{journal}{{Phys. Rev. Lett.}} \textbf{\bibinfo{volume}{{85}}},
  \bibinfo{pages}{{4643}} (\bibinfo{year}{{2000}}).

\end{thebibliography}
\end{document}